\documentclass[prd,aps,twocolumn,a4paper,floatfix]{revtex4}

\usepackage{graphicx,psfrag}
\usepackage{mathrsfs}
\usepackage{amsmath,amsfonts,amssymb}
\usepackage{multirow}

\def\p{\partial}

\def\non{\nonumber}                     
\def\half{\frac{1}{2}}
\def\bam{\textsc{BAM}}
\def\e{{\rm e}}
\def\i{{\rm i}}
\def\gccm{{\rm g\,cm^{-3}}}
\def\Msun{M_{\odot}}

\newcommand{\stext}[1]{\text{\tiny{#1}}}
\newcommand{\rhoadm}{\rho_{\stext{ADM}}}

\newcommand{\jIadm}[1]{S^#1_{~\stext{ADM}}}
\newcommand{\Sadm}{S_{\stext{ADM}}}
\newcommand{\Sijadm}[2]{S_{#1#2~\stext{ADM}}}

\newcommand{\TF}{\stext{TF}}

\usepackage{color}
\definecolor{cyan}{rgb}{0,0.9,0.9}
\definecolor{orange}{rgb}{0.9,0.5,0}
\definecolor{magenta}{rgb}{1,0,1}
\definecolor{purple}{rgb}{0.8,0.4,0.8}
\definecolor{gray}{rgb}{0.8242,0.8242,0.8242}
\begin{document}

\title{Numerical relativity simulations of binary neutron stars}

\author{Marcus \surname{Thierfelder}}
\email{marcus.thierfelder@uni-jena.de}
\author{Sebastiano \surname{Bernuzzi}}
\email{sebastiano.bernuzzi@uni-jena.de}
\author{Bernd \surname{Br\"ugmann}}
\email{bernd.bruegmann@uni-jena.de}
\affiliation{Theoretical Physics Institute, University of 
  Jena, 07743 Jena, Germany}

\date{\today}

\begin{abstract}
  We present a new numerical relativity code designed for simulations
  of compact binaries involving matter.
  The code is an upgrade of the {\bam} code to include general
  relativistic hydrodynamics and implements state-of-the-art 
  high-resolution-shock-capturing schemes on a hierarchy of mesh refined 
  Cartesian grids with moving boxes. 
  We test and validate the code in a series of standard experiments 
  involving single neutron star spacetimes.
  We present test evolutions of quasi-equilibrium
  equal-mass irrotational binary neutron star configurations in
  quasi-circular orbits which describe the late inspiral to merger phases. 
  Neutron star matter is modeled as a zero-temperature fluid; thermal
  effects can be included by means of a simple ideal-gas prescription.
  We analyze the impact that the use of different values of damping parameter 
  in the Gamma-driver shift condition has on the dynamics of the system.
  The use of different reconstruction schemes and 
  their impact in the post-merger dynamics is investigated.
  We compute and characterize the gravitational radiation emitted by
  the system. Self-convergence of the waves is tested, and we consistently estimate
  error-bars on the numerically generated waveforms in
  the inspiral phase.   
\end{abstract}

 
\maketitle


\section{Introduction}
\label{sec:intro}

Binary neutron stars (BNS) are among the most promising sources of
gravitational waves (GWs) for ground-based interferometers of present 
and future
generation~\cite{Belczynski:2006zi,Andersson:2009yt,:2010cfa}, and are also
at the origin of the powerful electromagnetic astrophysical
phenomena, in particular short-gamma-ray-bursts
(SGRBs)~\cite{Lee:2007js,Nakar:2007yr}. While other aspects of BNS 
(and neutron star, NS) physics are certainly interesting, these two
topics represent to date one of the most exciting observational and
theoretical challenge.  
On the one hand, the detection of GWs emitted during the last orbits of a
merger process is expected to convey unique information about the 
nature of matter at supra-nuclear densities which is largely 
unknown, e.g.~\cite{Read:2009yp}. On the other hand SGRBs are 
ultra-relativistic outflows likely to be injected during the post-merger
dynamics~\cite{Rosswog:2001fh,Rosswog:2003tn,Rezzolla:2011da}, but
neither a self-consistent model nor a simulation are yet available
to provide the conditions for the ``central
engine''~\cite{Paczynski:1986px,Narayan:1992iy}. 

A complete theoretical modeling of the late inspiral and merger phase
is possible only by means of numerical relativity (NR) simulations.  
While BNS simulations have a longer history, see
e.g.~\cite{Faber:2009,Duez:2009yz,Rosswog:2010ig} for recent reviews,  
the first NR simulation was performed in~\cite{Shibata:1999wm}, some
years before the first complete simulations of coalescing binary
black-holes (BBHs)~\cite{Pretorius:2005gq,Baker:2005vv,Campanelli:2005dd}. 
Nowadays a number of NR groups are performing BNS simulations~\cite{Font:1998hf,Font:2001ew,Miller:2003vc,Anderson:2007kz,Yamamoto:2008js,Liu:2008xy,Giacomazzo:2010bx}. 

Recent work investigated the evolution of irrotational, 
circularized, equal and unequal mass binaries without magnetic
fields~\cite{Kiuchi:2009jt,Yamamoto:2008js,Anderson:2007kz,Baiotti:2008ra,Rezzolla:2010fd}.  
In these works the dynamics of the system, mainly dependent on the
initial masses of the stars involved, is explored in detail with 
particular attention on the final product of the merger: either a
hyper-massive-NS (HMNS) (which eventually collapses on dynamical
timescales) or a black-hole (BH) with an accretion disk configuration
resulting from a prompt collapse. The gravitational radiation emitted by the
systems has been characterized.

Electromagnetic fields in NR simulations of BNS have been considered 
in~\cite{Liu:2008xy,Anderson:2008zp,Giacomazzo:2009mp,Giacomazzo:2010bx,Rezzolla:2011da}.
Their impact on GWs during the inspiral has been found negligible for
astrophysically motivated
intensities~\cite{Anderson:2008zp,Giacomazzo:2009mp,Giacomazzo:2010bx},
while they have certainly a major role in the post-merger phase  
for several astrophysical scenarios like SGRBs.  
Up to now electromagnetic fields have been considered in full general
relativity (GR) simulations only within the framework of ideal general
relativistic magneto-hydrodynamics (GRMHD), i.e.~in the limit of
infinite electrical conductivity. Some efforts towards resistive GRMHD
are ongoing~\cite{Palenzuela:2008sf}.    

  In all cases mentioned above the studies are based on the numerical
solution of the ideal general relativistic hydrodynamics
equations (GRHD), or GRMHD, coupled to some 3+1 hyperbolic
formulations
(BSSNOK~\cite{Nakamura:1987zz,Shibata:1995we,Baumgarte:1998te} or
GHG~\cite{Lindblom:2005qh}) of GR.   
Most of the NR results are obtained with simplified treatments of the 
NS interior, namely ideal gas, polytropic or piecewise polytropic
equations of state~(EoS). In~\cite{Shibata:2005ss} (and following
works) zero-temperature (cold) ``realistic'' EoS are also
employed. Thermal effects are expected to be relevant in the post-merger
phase. They are taken into account in an approximate way using hybrid
EoS~\cite{Shibata:2005ss,Bauswein:2010dn}, or in 
simulations based on approximations of GR which instead focus more on
microphysical
aspects (see e.g.~\cite{Ruffert:1995fs,Ruffert:1996by,Oechslin:2007gn}). 
The modeling of microphysics is particularly important in
combination electromagnetic fields in order to model SGRBs. 
Transport phenomena and neutrino
physics~\cite{Thorne:1981,Sekiguchi:2010ep,Sekiguchi:2010fh} are
currently not included in full GR BNS simulations (see however the
very recent work in~\cite{Sekiguchi:2011zd}). 

We also mention that mixed binaries, i.e.\ binary system composed of a 
black-hole and a neutron star, are currently under investigation with 
the same techniques employed for
BNSs~\cite{Shibata:2006ks,Loffler:2006nu,Etienne:2007jg,Etienne:2008re,Duez:2008rb,Duez:2009yy,Foucart:2010eq,Chawla:2010sw}.      

While some of the above mentioned aspects of BNS physics are quite
well understood, the complexity of the physical and technical problem 
certainly requires more investigations. On the one hand there is the need
of improving the microphysical modeling of the post-merger phase, on
the other hand there is the necessity to investigate the initial configuration
space and to improve the accuracy of the waveforms, cf.~\cite{Duez:2009yz}. 

The latter point is particularly important for future GW detection, see for example~\cite{Miller:2005qu}, 
also in view of a starting a program of collaboration between NR and
analytical relativity
community~\cite{Read:2009yp,Hinderer:2009ca,Damour:2009wj,Baiotti:2010xh,Baiotti:2011am}. 
In contrast to BBH simulations, where a precise waveform
analysis is routinely performed (see
e.g.~\cite{Husa:2007hp,Hannam:2010ec}), the accuracy of NR  
waveform from BNS has been poorly investigated so far. 
Ref.~\cite{Baiotti:2009gk} reports a first error analysis of BNS
waveforms. The curvature ($\psi_4$) waveforms are found to converge at
order 1.8 in the inspiral phase and at 1.2 after the
merger. The waveforms are aligned to perform the convergence
test~\cite{Pollney:2007ss}. An error budget regarding the grid 
setting is also presented. 
In~\cite{Baiotti:2011am} a similar error budget regarding finite
extraction, thermal and resolution effects is presented.
An estimate of the phase and amplitude errors during inspiral is given
based on a convergence rate assumed from other
simulations~\cite{Baiotti:2009gk}.
The main source of error has been found to be the finite resolution.
We are not aware of other detailed investigations regarding
convergence and precise error estimates on the numerically 
generated GWs.

In this paper we present a new code to simulate non-vacuum spacetimes
in full GR. The code is an extension of the {\bam} 
code developed at Jena and elsewhere \cite{Brugmann:2008zz,Bruegmann:2003aw,Bruegmann:1997uc}
for numerical studies of multiple black holes
spacetimes with adaptive mesh refinement techniques. The {\bam} code has
been upgraded in order to solve the flux-conservative Eulerian formulation
of ideal GRHD equations~\cite{Banyuls:1997zz} coupled to the Einstein
system. We describe in detail the equations and the implementation of
HRSC scheme in {\bam}. The code allows the use of hybrid EoS composed of
a cold part, generically provided by tables, and a thermal part
modeled with an ideal gas EoS~\cite{Shibata:2005ss,Bauswein:2010dn}.
A simple thermodynamical consistent procedure for table interpolation is 
employed.

We validate the code against a number of stringent tests involving 
single-star spacetimes. Each test permits the verification of a part of
the numerical algorithm. Convergence and constraint violation in
particular are discussed in some detail considering different
reconstruction methods. 

We present our first results concerning the simulation of
gravitational radiation emitted from BNS evolutions. 
Since we do not simulate magnetic fields, our main interest is
related to the GWs emitted during the inspiral phase. 
We focus, as a test case, on a binary already considered in the
literature~\cite{Taniguchi:2002ns,Baiotti:2008ra,Giacomazzo:2010bx}. 
The initial irrotational configuration in quasi-circular orbit has been 
evolved for about three orbits, the merger and the 
post-merger phase with both a cold and a hot EoS. 
We investigate the effect of the use of different
reconstruction methods on the dynamics.
We report new results concerning the role of the damping parameter, $\eta$, 
in the Gamma-driver shift condition with regard to the dynamics.
Waveforms are extracted from the simulation via a standard algorithm
based on the Newman-Penrose scalar $\psi_4$. 
The actual GWs (metric waveforms) are additionally reconstructed with 
two different methods, namely a time-domain~\cite{Damour:2008te,Baiotti:2008nf} and 
a spectral~\cite{Reisswig:2010di} integration. The two methods  are
compared. We discuss the convergence of the numerically extracted GWs.
Using different convergence series we provide the first consistent
estimate of the errors on phase and amplitude in BNS simulations. 
This work is our first contribution to the study of BNS spacetime by NR
simulations; the method presented here will serve as a future
reference.

The structure of the paper is as follows. 
In Sec.~\ref{sec:eqs} we review the basic equations solved in {\bam}.
In Sec.~\ref{sec:eos} we review the description of the NS matter within BNS simulations in NR.
In Sec.~\ref{sec:numerics} we summarize the numerical method adopted for solving the GRHD 
equations as well as the singularity and vacuum treatment in {\bam}.
In Sec.~\ref{sec:ns} we present our results concerning single star spacetimes.
In Sec.~\ref{sec:bns} we present our results on simulations of BNSs.

Dimensionless units $c=G=\Msun=1$ are employed. 
Times and lengths are often expressed in $ms$ and $km$ to facilitate
comparison with the literature. Indexes $a,b,c,...$ run from $0$ to
$3$, indexes $i,j,k,...$ from $1$ to $3$.


\section{Equations}
\label{sec:eqs}

In this section we review the equations solved in {\bam}. 
We assume the usual 3+1 decomposition of spacetime; 
the metric reads
\begin{equation}
ds^2 = -(\alpha^2 - \beta^i \beta_i) dt^2 + 2\beta_i dx^i dt +\gamma_{ij}dx^ix^j \ ,
\end{equation}
where $\alpha$ and $\beta^i$ are the lapse and shift vector and
$\gamma_{ij}$ the spatial 3-metric. The Einstein equations are formulated
in the strongly hyperbolic BSSNOK form and presented in
Sec.~\ref{sec:einstein}. While the BSSNOK system is described in several
textbooks we describe it again in the following  for completeness. The
equation are explicitly written for the $\chi$-BSSNOK system not given
in~\cite{Brugmann:2008zz}; we point out some relevant detail for the
implementation and correct some minor typos.
GRHD equations are given in Sec.~\ref{sec:grhd} following the
flux-conservative formulation of~\cite{Banyuls:1997zz}.

\subsection{Metric}
\label{sec:einstein}

The BSSNOK formalism assumes the conformal decomposition of the 3-metric,
\begin{equation}
    \tilde \gamma_{ij} = e^{-4\phi} \gamma_{ij}
    \quad\textrm{and}\quad 
    \phi = \frac{1}{12} \ln \det \gamma_{ij} \ ,
\end{equation}
where $\tilde \gamma_{ij} $ is the conformal 3-metric and $\phi$ the conformal factor. 
Following~\cite{Campanelli:2005dd} we introduce the variable
\begin{equation}
    \chi \equiv e^{-4\phi} \ .
\end{equation}
The extrinsic curvature $K_{ij}$ of the spatial hypersurfaces is decomposed as
\begin{equation}
    \tilde A_{ij} = \chi \left(K_{ij}-\frac{1}{3} \gamma_{ij} K\right) 
    \quad\textrm{and}\quad
    K \equiv \gamma^{ij} K_{ij} \ .
\end{equation}
The evolution system reads
\begin{widetext}
\begin{eqnarray}
(\partial_t - \mathcal{L}_\beta) \chi &=& \frac{2}{3}\alpha \chi K \ , \\
\label{eq:evo:bssn1}
(\partial_t - \mathcal{L}_\beta) \widetilde{\gamma}_{ij} &=& -2 \alpha \widetilde{A}_{ij} \ , \\
\label{eq:evo:bssn2}
(\partial_t - \mathcal{L}_\beta) \widetilde{A}_{ij} &=& 
    \chi\left[ -D_i D_j \alpha + 
    \alpha ( R_{ij} - 8\pi \Sijadm{i}{j})\right]^{\TF} 
    - \frac{1}{3}\widetilde{\gamma}_{ij}\alpha (16\pi\rhoadm) 
    + \alpha\left( K \widetilde{A}_{ij} -2 \widetilde{A}_{ik} {\widetilde{A}^k}_j \right) \ , \\
\label{eq:evo:bssn3}
(\partial_t - \mathcal{L}_\beta) K &=& -D^i D_i \alpha +
    \alpha \left( \widetilde{A}_{ij} \widetilde{A}^{ij} + \frac{1}{3}K^2 \right) 
    + 4\pi\alpha (\rhoadm + \Sadm) \ , \\
\label{eq:evo:bssn4}
\partial_t \widetilde{\Gamma}^i &=& \widetilde{\gamma}^{ik}\partial_j\partial_k\beta^j 
    + \frac{1}{3}\widetilde{\gamma}^{ij}\partial_j\partial_k\beta^k
    + \beta^j\partial_j\widetilde{\Gamma}^i  
    - \widetilde{\Gamma}^j\partial_j\beta^i 
    + \frac{2}{3}\widetilde{\Gamma}^i\partial_j\beta^j
    - 2\widetilde{A}^{ij}\partial_j\alpha  
    \\ & & 
    + 2\alpha \left( \widetilde{\Gamma}^i_{jk}\widetilde{A}^{jk} - \frac{3}{2\chi}\widetilde{A}^{ij} \partial_j\chi
    - \frac{2}{3}\widetilde{\gamma}^{ij} \partial_j K  \right. 
     \left. -\frac{8\pi}{\chi}\jIadm{i} \right) \nonumber\ .
\label{eq:evo:bssn5}
\end{eqnarray}
\end{widetext}
Above $\mathcal{L}_\beta$ is the Lie derivative along the shift
vector, $D_i$ the covariant derivative associated with $\gamma_{ij}$, 
$R_{ij}$ is the Ricci tensor, see Eq.\ (18)-(20) in~\cite{Brugmann:2008zz}, 
and the York-ADM quantities are defined in Sec.~\ref{sec:grhd}.
The terms proportional to $\rho_{\rm ADM}$ in
Eqs.~\eqref{eq:evo:bssn2} and \eqref{eq:evo:bssn3} appear because the
trace of $R_{ij}$ is obtained by substituting the Hamiltonian
constraint as usual, and in Eq.~\eqref{eq:evo:bssn2} we choose to use the
vacuum constraint in $(R_{ij})^{\TF}$ and to add the missing term
separately.
The variables
\begin{equation}
\widetilde{\Gamma}^i \equiv \tilde{\gamma}^{jk} \tilde{\Gamma}^i_{jk} \ ,
\end{equation}
defined in term of the Christoffel symbols of the conformal metric, 
are promoted to new evolution variables.

The constraints are
\begin{eqnarray}
  G_i     &\equiv& \tilde{\gamma}_{ij}\tilde{\Gamma}^{j}-
  \tilde{\gamma}^{jk} \partial_k \tilde{\gamma}_{ij} \ , \\
  H       &\equiv& R -\tilde{A}^{ij}\tilde{A}_{ij}+\frac{2}{3}
  K^2-16\pi\rho_{\rm  ADM}\ ,\\
  M^i &\equiv& \p_j\tilde{A}^{ij} + \tilde{\Gamma}^i{}_{jk}\tilde{A}^{jk}
  -\frac{2}{3}\tilde{\gamma}^{ij}\p_j \hat{K} 
  -\frac{3}{2}\tilde{A}^{ij} \partial_j \log\chi \nonumber\ .\\
\end{eqnarray}

The gauge is specified by the 1+log lapse and Gamma-driver-shift~\cite{Bona:1994b,Alcubierre:2002kk,vanMeter:2006vi,Gundlach:2006tw}:
\begin{align}
\label{eq:lapse_1log}
\p_t\alpha   &= \beta^i\partial_j\alpha_{i}-\alpha^2\mu_LK \ ,\\
\label{eq:beta_Gdriv}
\p_t\beta^i  &= \mu_S\tilde{\Gamma}^i-\eta
\beta^i+\beta^j\partial_j\beta^i \ .
\end{align}
The parameters are fixed to $\mu_S=3/4$ and $\eta=0.3$ if
not stated otherwise.

\subsection{Matter}
\label{sec:grhd}

We assume matter composed of a single particle species (simple
  fluid) and described by the perfect fluid stress-energy tensor
\begin{equation}
  \label{eq:tmunu}
  T_{ab}=\rho h u_a u_b + p g_{ab} \ ,
\end{equation}
where $\rho$ is the rest-mass density,
$\epsilon$ is the specific internal energy,
$h\equiv1+\epsilon + p/\rho$ is the specific enthalpy,
$p$ is the pressure, 
and $u^a$ is the 4-velocity ($u^a u_a=-1$) of the fluid.
The total energy density is given by
$\e=\rho(1+\epsilon)$.

The GRHD equations for the perfect fluid matter (ideal GRHD)
are the local conservation law for the energy-momentum tensor, the
conservation law for the baryon number, and the equation of state of
the fluid: 
\begin{eqnarray}
  \nabla_a T^{ab} & =& 0 \ , \label{eq:divT}\\
  \nabla_a \left( \rho u^a \right) & = & 0 \ , \label{eq:rhoua}\\
  P(\rho,\epsilon) &=& p  \label{eq:eos} \ .
\end{eqnarray}
Following~\cite{Banyuls:1997zz} we rewrite
Eqs.~\eqref{eq:divT}-\eqref{eq:rhoua}  
in first-order flux-conservative form,
\begin{eqnarray}
  \label{eq:hydro_consform}
  \partial_t \vec{q} + \partial_i \vec{f}^{(i)} (\vec{q}) = \vec{s}(\vec{q}) \ ,
\end{eqnarray}
by introducing the \emph{conservative} variables
\begin{eqnarray}
  \vec{q} &=& \sqrt{\gamma}\{ \, D, \, S_k, \, \tau \, \} \ ,
\end{eqnarray} 
where 
\begin{eqnarray}  
  D &\equiv& W\rho \ , \non \\
  S_k &\equiv& W^2 \rho h  v_k \ , \label{eq:hydro_cons} \\
  \tau &\equiv& \left(W^2 \rho h - p\right) - D \non \ .
\end{eqnarray}
The simple physical interpretation of these variables is that they
represent the rest-mass density ($D$), the momentum density ($S_k$)
and an internal energy ($\tau=\rho_{\rm ADM}-D$) as viewed by Eulerian
observers. Above $v^i$ is the fluid velocity measured by the Eulerian observer with
\begin{equation}
  v^i = \frac{u^i}{W} + \frac{\beta^i}{\alpha} = 
  \frac{1}{\alpha}\left( \frac{u^i}{u^0}+ \beta^i \right) \ .
\end{equation}
$W$ is the Lorentz factor between the fluid frame and the Eulerian observer, 
$W=1/\sqrt{1-v^2}$, with $v^2=\gamma_{ij}v^i v^j$.
The fluxes in Eq.~(\ref{eq:hydro_consform}) are
\begin{widetext}
  \begin{equation}
    \label{eq:hydro_fluxes}
    \vec{f}^{(i)} = \sqrt{-g}\left\{ D\left(v^i-\frac{\beta^i}{\alpha}\right) \ , 
    S_k\left(v^i-\frac{\beta^i}{\alpha}\right) + p\delta_k^i \ , 
    \tau\left(v^i-\frac{\beta^i}{\alpha}\right)+pv^i \right\} \ ,
  \end{equation}
while the source terms are
  \begin{align}
    \label{eq:hydro_sources}
    \vec{s} &= \sqrt{-g}\left\{0,
    T^{ab}\left( \partial_a g_{a k} - \Gamma^\delta_{ab}g_{\delta k}\right),
  \alpha\left( T^{a 0}\partial_a \ln \alpha - T^{ab} \Gamma^0_{ab} \right) \right\} \\
  %
  %
  \label{eq:hydro_sources_sder}  
  & = \sqrt{-g}\left\{0, 
  T^{00}\left (\half \beta^i\beta^j \partial_k \gamma_{ij} - \alpha \partial_k \alpha \right)  + 
  T^{0i}\beta^j\partial_k \gamma_{ij}  + T^0_i\partial_k\beta^i + \half T^{ij}\partial_k \gamma_{ij} ,\right. \non \\
  & \qquad  T^{00}\left( \beta^i\beta^j K_{ij} - \beta^i\partial_i \alpha\right)
  + T^{0i}\left( 2\beta^j K_{ij} - \partial_i \alpha \right)  +
  T^{ij}K_{ij} \biggr\} \ .
\end{align} 
Above $g\equiv \det g_{ab}=-\alpha^2\gamma$ with 
$\gamma\equiv \det \gamma_{ij}$. From 
straightforward calculations the stress-energy tensor is
\end{widetext}
\begin{eqnarray}
  \label{eq:tmunu_comp}
  T^{00} &=& \frac{\rho h W^2 - p}{\alpha^2} \ , \\
  T^{0i} &=& \frac{\rho h W^2 (v^i-\frac{\beta^i}{\alpha})}{\alpha} + \frac{p\beta^i}{\alpha^2} \ , \\
  T^{ij} &=& \rho h W^2
  (v^i-\frac{\beta^i}{\alpha})(v^j-\frac{\beta^j}{\alpha}) + p(\gamma^{ij} -
  \frac{\beta^i\beta^j}{\alpha^2}) \ , \\
  T^0_i &=&  \frac{\rho h W^2}{\alpha} v_i \ .
\end{eqnarray}
Note that both the fluxes and the source terms depend also on the
\emph{primitive} variables 
$\vec{w}=\{p, \rho, \epsilon, v^i \}$, and the source terms 
do not depend on derivatives of $T_{ab}$. Eq.~\eqref{eq:hydro_consform}  above 
conserves exactly the rest-mass (or baryonic mass),
\begin{equation}
\label{eq:M0}
M_0 \equiv \int d^3x~q^0 = \int d^3x~\sqrt{\gamma}D \ .
\end{equation} 
Standard York-ADM matter variables are easily recovered
\begin{eqnarray}
  \rho_{\rm ADM} & \equiv & n^a n^b T_{ab} = \rho h W^2 - p = \tau + D \ , \\
  S^i_{\rm ADM}&\equiv &  -n^a\gamma^{ib}T_{ab} = \rho h W^2 v^i = S^i \ , \\
  S^{ij}_{\rm ADM}&\equiv &  \gamma^{ia}\gamma^{jb}T_{ab} = \rho h W^2 v^i v^j + \gamma^{ij}p \ .
\end{eqnarray}
The system in Eq.~\eqref{eq:hydro_consform} is strongly hyperbolic 
provided that the EoS is causal (the sound speed is less than the speed 
of light)~\cite{Banyuls:1997zz}. Eigenvalues (in direction $x$)  are given by
\begin{eqnarray}
  \label{eq:eigenv}
  \lambda_0 &=& \alpha v^x - \beta^x  \ , \\
  \label{eq:eigenv1}
  \lambda_\pm &=& \frac{\alpha}{1-v^2c_s^2}
  \left[ 
    v^x\left( 1-c_s^2 \right) \pm 
    \right. \\
    & &\left. 
    c_s \sqrt{ (1-v^2)[\gamma^{xx}( 1-v^2c_s^2) 
        - v^xv^x( 1-c_s^2 )] } 
    \right] - \beta^x \; .\nonumber
\end{eqnarray}
The others are obtained by permutation of indexes.
Above the sound speed is defined by
\begin{align}
\label{eq:cs2}
c^2_s = \left( \chi + \frac{P}{\rho^2} \kappa\right)\frac{1}{h} \ , \\
\chi \equiv \frac{\partial P}{\partial \rho} \ \ \ , \ \ \  
\kappa \equiv \frac{\partial P}{\partial \epsilon} \ .
\end{align}
Note the notation conflict between $\chi$ and the metric variable defined in Sec.~\ref{sec:einstein}, 
which is resolved in the following by the context of the discussion.
In the Sec.~\ref{sec:eos} we discuss the modeling of NS interior,
i.e.\ the EoS of the fluid.


\section{Description of NS matter}
\label{sec:eos}

The exact nature of the internal structure of a NS is unknown. 
The standard picture assumes that the matter of an isolated NS in
hydrostatic equilibrium is strongly degenerate and at thermodynamic
equilibrium. Consequently temperature effects are 
neglected and the matter is in its ground state: cold catalyzed
  matter. Under these conditions the EoS has  
one-parameter character~\cite{Shapiro:1983du}, $p=P(\rho)$. 
Note that if one has an one-parameter EoS the GRHD equation
for $\tau$ is equivalent to that for $D$ and thus redundant. 
A simple cold EoS often employed in NR simulations is the polytropic EoS
\begin{equation}
\label{eq:eos:poly}
P(\rho) = K \rho^\Gamma \ ,
\end{equation}
where $K$ and $\Gamma$ are parameters. The polytropic EoS describes 
isentropic fluids and it is equivalent to the well known ideal gas 
\begin{equation}
\label{eq:eos:ideal}
P(\rho,\epsilon) = \left( \Gamma-1\right) \rho \epsilon \ ,
\end{equation}
if the flow remains smooth, i.e.\ no shock-heating. 
The parameter for a relativistic Fermi gas is $\Gamma=4/3$, and 
$1<\Gamma<3$ for NS simulations.
Note that it coincides with the adiabatic index of the fluid,
\begin{equation}
\Gamma\equiv c^2_s \left( \frac{\e}{P} +1 \right) \ .
\end{equation}
The constant $K$ in Eq.~\eqref{eq:eos:poly} fixes the entropy.
Ideal gas and polytropic EoSs provide a simple and analytical
description of NS matter, even though it is quite rough and approximate.

The NS structure consists qualitatively~\cite{Haensel:2007yy} of an
outer region (outer crust) that extends until the neutron drip
$\e_{\rm drip}\sim10^{11}$ $\gccm$, an inner crust up to nuclear densities,
$\e_{\rm nuc}\gtrsim10^{14}$ $\gccm$, which characterize the central core.  
The composition of the inner and outer crust is reasonably well
understood. Matter in the outer crust consists of a Coulomb lattice
immersed in an electron gas; the pressure is dominated by electron
pressure. As density increases, the pressure contribution from neutrons 
becomes larger, and in the inner crust the EoS softens due to the attractive
long-range behavior of the strong interaction.
Models for the outer and inner crust are for example the
BPS~\cite{Baym:1971a} and the BBP~\cite{Baym:1971b}, or
HP94~\cite{Haensel:1994} EoSs, respectively. At densities $\e>\e_{\rm nuc}$ nuclei
are not stable and a plasma of nucleons dominates the pressure. The EoS
stiffens due to the repulsive short-range character of strong
interactions. The modeling of the matter in the core is difficult
and requires assumptions on the nucleon-nucleon potential and the
solution of the many-body problem. Further complications are the
presence of hyperons and the necessity to solve the relativistic
many-body problem, super-fluidity, pion condensation, and phase
transition to quark matter.

The NS core contains most of the mass (98\%), while 
only a very small fraction is in the outer crust ($10^{-5}$ \%). Despite
this the (outer and inner) crust plays the major role in defining
tidal the deformation-disruption (mass-shedding limit) during evolutionary
scenarios. 
Several EoS for the core are proposed in the literature, see
e.g.~\cite{Arnett:1976dh,Stergioulas:1994ea}.  
We refer to the set of these EoSs generically as \emph{realistic}
EoS. They are provided by tables, or, alternatively, 
they can be phenomenologically parametrized by piecewise polytropes~\cite{Read:2008iy}. 
Most of the realistic EoSs are considered ``equivalent'' 
since they are able to produce NSs with mass and radii that agree
with observational constraints. See e.g.~\cite{Lattimer:2006xb,Ozel:2009da,Steiner:2010fz} 
and references therein for a detailed report on observational constraints on NS
EoSs. Gravitational waves emission from pulsating NSs 
represents a unique opportunity to constrain the EoS of the
core~\cite{Andersson:1997rn}. 
 
In a dynamical scenario (e.g.\ merger or collapse) NS matter can be
heated, thus acquiring a thermal component. 
This situation can not be modeled if cold (one-parameter) EoS are
employed and the use of a \emph{hot} (temperature dependent) EoS is
necessary. The simplest hot EoS is the ideal gas
Eq.~\eqref{eq:eos:ideal}. A simple way to extend 
nuclear cold EoS to include a hot component is 
presented in~\cite{Shibata:2005ss}. Given values of $\rho$ and
$\epsilon$, a hot part is allowed evolving also the $\tau$ equation
and defining a ``hot internal energy'' as the difference between the
actual $\epsilon$ and the $\epsilon^{\rm cold}$ from the realistic
cold EoS,
\begin{equation}
\epsilon^{\rm hot} \equiv \epsilon -\epsilon^{\rm cold} \ .
\end{equation}
The pressure is augmented with a thermal component which is modeled via a
simple ideal gas EoS,
\begin{eqnarray}
  \label{eq:eos_tab1Dhot}
  P(\rho,\epsilon) &=& P^{\rm cold}(\rho) + P^{\rm hot}(\rho,\epsilon) \\
  &=& P^{\rm cold}(\rho) + (\Gamma-1)\rho\epsilon^{\rm hot} \ .
\end{eqnarray}
The following relations hold:
\begin{eqnarray}
  \label{eq:eos_chi}
  \chi &=& \frac{2P^{\rm cold}}{\rho} 
  + \rho^2\frac{d^2\epsilon^{\rm cold}}{d\rho^2} \\
  && + (\Gamma-1)\epsilon^{\rm hot} - 
    (\Gamma-1)\rho\frac{d\epsilon^{\rm cold}}{d\rho} \ , \non\\
  \kappa &=& (\Gamma-1)\rho \ .
\end{eqnarray}
Note that the adiabatic index in this case does not coincides with the
sum of the ``cold'' adiabatic index and the ideal gas $\Gamma$. 
Hot EoSs constructed in this way are called \emph{hybrid}.
A study of the validity of this approach can be found in~\cite{Bauswein:2010dn}. 
Note that obviously the hybridization of the polytropic EoS is the
ideal gas EoS.

The only complete (including temperature dependence) microphysical
EoSs have been developed by Shen et al.~\cite{Shen:1998gq}, Lattimer and
Swesty~\cite{Lattimer:1991nc}, and recently by
Shen~G. et al~\cite{Shen:2011kr,Shen:2011fc}. They allow 
inclusion of neutrino emission schemes, e.g.~\cite{O'Connor:2009vw},
and currently provide the best model to describe high density NS matter.

In addition to the analytic ones, our code can handle cold EoSs provided
by tables and implements the hybridization procedure described
above. While the code is ready to host hot complete EoS, we do not
consider them in this work and postpone their use to the future.


\section{Numerical method}
\label{sec:numerics}

Our code is part of the {\bam}
code~\cite{Brugmann:2008zz,Bruegmann:2003aw,Bruegmann:1997uc}, 
extending it with a module for GRHD. In the following we focus only 
on the matter solver, referring to~\cite{Brugmann:2008zz}   
for a description of the code infrastructure and the algorithm for the 
solution of the Einstein equations via the BSSNOK scheme.
We mention only that the evolution algorithm is based on the
method of lines (MoL) with explicit Runge-Kutta methods (3rd order in this work) and finite
differences in space (4th order in this work). Mesh refinement is provided 
by a hierarchy of cell-centered nested Cartesian grids and Berger-Oliger time 
stepping. Metric variables are interpolated in space by means of 4th order 
Lagrangian polynomials and matter conservatives by a 4th order WENO 
scheme (see below for details about the latter).
Interpolation in Berger-Oliger time stepping is performed at 2nd order. 
Some of the mesh refinement levels can be dynamically moved and
adapted during the time evolution according to the technique of
``moving boxes''. GWs are extracted using the $\psi_4$ formalism (see
Sec.~III of~\cite{Brugmann:2008zz}).

The algorithm implemented for the matter is a robust
high-resolution-shock-capturing (HRSC)
method~\cite{Font:2007zz,Marti:1999wi} based on a central scheme for the 
numerical fluxes. It has been successfully used and tested in
spherical symmetry in~\cite{Bernuzzi:2009ex}. HRSC schemes represent nowadays
the state-of-the-art methods to solve GRHD equations since they can
properly handle physical shocks, steep gradients and high Lorentz
factors in relativistic plasma. In the following we describe in detail
our implementation as well as the implementation of the EoS interface
and the treatment of vacuum regions and spacetime singularities.

\subsection{HRSC scheme for matter}
\label{sbsec:hrsc}

The GRHD equations are solved by means of a HRSC
method~\cite{Font:2007zz,Marti:1999wi} which considers the
semi-discrete form of the equations and combines the Runge-Kutta
integration with a cell-centered scheme for the RHS based on robust
central schemes or simple Riemann solvers~\cite{Kurganov:2000}.  
Both the time stepping and the spatial refined mesh are shared with
the metric system.  
Our implementation follows quite closely the algorithm presented
in~\cite{DelZanna:2002rv} (see
also~\cite{DelZanna:2007pk,Kurganov:2000,Nessyahu:1990,Shu:1988,Shu:1989}).  
The semi-discrete form of Eq.~\eqref{eq:hydro_consform} is
\begin{equation}
  \label{eq:grhd_semid}
  \frac{d\vec{q}_{i,j,k}}{dt} = \vec{s}_i + \frac{1}{h}
  \left( 
  \vec{F}_{i-\frac{1}{2},j,k} - \vec{F}_{i+\frac{1}{2},j,k} 
  \right) + {\rm other\ directions} \ ,
\end{equation} 
where $h$ is the grid spacing and
$\vec{F}_{i\pm\frac{1}{2},j,k}$ are the numerical fluxes (both in the $x$
direction). For simplicity the description is limited to 1D
since all the steps necessary to construct the numerical fluxes can be 
(and actually are) performed in one direction at a time.
In Eq.~\eqref{eq:grhd_semid} the difference of the numerical
fluxes is the Taylor approximation, at a certain order $r$, for the
divergence of the fluxes
\begin{eqnarray}
  \label{eq:div_approx}
  h \vec{f}'(x_i)  &=& \left( \vec{F}_{i+\frac{1}{2}} - \vec{F}_{i-\frac{1}{2}} \right) +\mathcal{O}(h ^r) \ ,\\
  \label{eq:num_flx}
  \vec{F}_{i+\half} &\equiv& \vec{f}_{i+\half} +\sum_{j=1}^{(r-1)/2} c_{2j} D^{(2j)}\vec{f}_{i+\half} \ ,
\end{eqnarray}
where the interface fluxes $\vec{f}_{i+\half}$ are computed with a
Riemann solver and the $D^{(2j)}$ is a discrete operator which
approximates derivatives of order $2j$.
We consider only 2nd order schemes, which amounts to dropping
high-order terms in Eq.~\eqref{eq:num_flx}. 
The interface fluxes are computed by the local Lax-Friedrichs (LLF)
central scheme~\cite{Nessyahu:1990,Kurganov:2000},  
or the  well known two-speed HLL~\cite{Harten:1983} Riemann solver.
In this work we focus on LLF which, in spite of its simplicity, 
has been proved to be robust and competitive with respect to approximate
Riemann solvers~\cite{LucasSerrano:2004aq} also in the case of neutron
star simulations~\cite{Shibata:2005jv}. The LLF flux reads
\begin{equation}
  \label{eq:flux_llf}
    \vec{f}^{~\rm (LLF)}_{i+\half} = \half\left[ \vec{f}^L + \vec{f}^R - a \left( \vec{q}^L - \vec{q}^R\right) \right] \ .
\end{equation}
The parameter $a$ is the maximum of the local characteristic speeds
(eigenvalues computed at interfaces $i\pm\half$)  of the system and it
is the only characteristic information used.  
The quantities written with superscript $L/R$ are the physical fluxes
(Eq.~\eqref{eq:hydro_fluxes}) and the conservative variables computed
(``reconstructed'') at the interface $i+\half$.
The reconstruction is performed using a non-oscillatory interpolation
centered, respectively, on $i$ ($L$, left) and on $i+1$ ($R$,
right). The reconstruction step is performed on the primitive
variables. Several methods are implemented in the code: linear Total
Variation Diminishing (TVD) interpolation based on ``minmod'' (MM2) and
``monotonized centered'' (MC2) slope limiters (see
e.g.~\cite{Toro:1999} and~\cite{Font:1999wh}),  the interpolation of
the piecewise parabolic method
(PPM)~\cite{Colella:1982ee,Marti:1994,Mignone:2005ns}, and the third
order convex-essentially-non-oscillatory (CENO3) algorithm
by~\cite{Liu:1998,Zanna:2002qr}.  
The construction of the numerical fluxes requires an interpolation of
metric quantities at the interfaces.  
We implemented both 2nd order (simple averages) and 4th order
Lagrangian interpolation. The latter is the default used in all 
the tests presented here. 

The algorithm used to recover the primitive variables depends on the
choice of the EoS. In case of a general EoS we adopt the standard
algorithm  described in~\cite{Marti:1999wi}, that employs the EoS of
the form Eq.~\eqref{eq:eos} and a Newton-Raphson method. 
In case of cold one-parameter EoS a similar procedure is adopted but
based on the equation which defines the variable
$D$~\cite{Baiotti:2004wn}. The exact equations employed as well as the
methods are detailed in Appendix~\ref{app:c2p}.

Boundary conditions are applied on the primitive variables before the
reconstruction step by simple extrapolation (``outflow'').

We describe now the spatial interpolation used for the
conservative variables in the mesh refinement, i.e.\ between levels. 
A non-oscillatory interpolation is necessary in order to avoid the Gibbs
phenomenon. Different methods are adopted in different codes, see 
e.g.~\cite{Neilsen:2005rq,Baiotti:2010ka}.   
As anticipated at the beginning of the section we adopt a 4th order
WENO algorithm as described in~\cite{Macdonald:2008}. The 1D scheme is 
summarized in the following. Given the four points $x_{i-1}, x_i,
x_{i+1}, x_{i+2}$ and the corresponding data $f_{i-1}, f_i, f_{i+1},
f_{i+2}$, two candidate interpolating polynomials are constructed as
\begin{eqnarray}
    p_1(x) &=& f_i + \frac{f_{i+1}-f_{i-1}}{2 h}(x-x_i) + \nonumber\\
    && \frac{f_{i+1}-2f_{i}+f_{i-1}}{2 h^2}(x-x_i)^2 \ , \nonumber\\
    p_2(x) &=& f_i + \frac{-f_{i+2}+4f_{i+1}-3f_i}{2 h}(x-x_i) + \nonumber\\
    && \frac{f_{i+2}-2f_{i+1}+f_{i}}{2 h^2}(x-x_i)^2 \ .
\end{eqnarray}
The final interpolated value is given by
\begin{equation}
    p(x) = w_1(x) p_1(x) + w_2(x) p_2(x) \ ,
\end{equation}
where the weights are 
\begin{eqnarray}
  w_i &=& \frac{\alpha_i(x)}{\alpha_1(x)+\alpha_2(x)} \ ,\\
  \alpha_i &=& \frac{C_i(x)}{(\varepsilon+\textrm{IS}_i)^2} \ .
\end{eqnarray}
The weights are defined in term of the smoothness indicators 
\begin{eqnarray}
  {\rm IS}_1 &=& \frac{25}{12}f_{i+1}^2 + \frac{64}{12}f_i^2 + \frac{13}{12}f_{i-1}^2 + \nonumber\\
  &&  \frac{26}{12}f_{i+1}f_{i-1} - \frac{52}{12}f_i f_{i-1} - \frac{76}{12}f_{i+1} f_i \ , \nonumber\\
  {\rm IS}_2 &=& \frac{25}{12}f_i^2 + \frac{64}{12}f_{i+1}^2 + \frac{13}{12}f_{i+2}^2 + \nonumber\\
  &&  \frac{26}{12}f_{i+2} f_i - \frac{52}{12}f_{i+2} f_{i+1} -
  \frac{76}{12}f_{i+1} f_i \ ,
\end{eqnarray}
and the optimal weights
\begin{eqnarray}
  C_1(x)&=&\frac{x_{i+2}-x}{3 h} \ , \\
  C_2(x)&=&\frac{x-x_{i-1}}{3 h} \ .
\end{eqnarray}
In the case of a smooth function, the interpolation reduces to 
standard 4th order Lagrangian interpolation. For less regular
functions the order of interpolation drops based on the local
continuity of the derivatives (see discussions in
e.g.~\cite{Jiang:1996,Tchekhovskoy:2007zn}). 
In the implementation we set $\varepsilon=10^{-6}$ to avoid division 
by zero.
The algorithm has been slightly modified in order to 
enforce monotonicity in the solution~\cite{Yamamoto:2008js}.
If $p_i(x)$ at a given point is larger or smaller than all
four function values $f_i$, we set the corresponding $\alpha_i$ to zero.
If all $\alpha_i$ are zero we use linear interpolation. 

Finally we comment about the overall convergence rate expected in the simulation data.
The different elements of the algorithm described above contribute differently to the error budget, 
some errors converge away more rapidly while others dominate. 
According to our previous experience with black hole spacetimes and to other results in the literature, 
we expect the overall error to be dominated by the truncation error of the finite difference discretization of the right-hand-side.
In this case the matter HRSC scheme is 2nd order accurate at most, thus the evolved fields and the relevant post-processed quantities,  
such as the gravitational waves and the constraints, are expected to converge to the continuum solution at this rate.

\subsection{Equation of state}
\label{sbsec:eos}

Polytropic and ideal gas EoSs are analytical and do not require special
treatment. However the use of 
realistic EoSs, provided in form of tables, poses the problem of
interpolation. A major requirement is that the interpolation is
consistent with thermodynamics. The application of the first law at
zero temperature translates into the relation
\begin{equation}
  \label{eq:1stlaw}
  P(\rho) = \rho^2\frac{d\epsilon}{d\rho} \ .
\end{equation}
Several methods are adopted in the literature. Widely
used is a simple and efficient linear interpolation, see
e.g.~\cite{Corvino:2010yj}, which is not thermodynamically consistent.
Hermite polynomials can be adopted for a thermodynamically consistent
interpolation of general EoSs~\cite{Swesty:1996}. 
After~\cite{Haensel:2004nu} analytic fits of the tables
were also used in simulations~\cite{Shibata:2005ss} in a
thermodynamical consistent way. 

In our code we adopt the following thermodynamically
consistent procedure for cold EoS. 
In GRHD evolutions the quantities provided by
the EoS are $p$, $\chi$ and $\kappa$. They are obtained in two
steps. First, for a given $\rho$ we construct $\epsilon(\rho)$ and its
derivatives by interpolating the logarithms, i.e. the function $y(x)$
with $y=\log_{10}\epsilon$ and $x=\log_{10}\rho$.
Derivatives are taken consistently from the interpolating polynomial.
Second, the pressure is obtained from Eq.~\eqref{eq:1stlaw} together with 
$\chi$, which requires $\epsilon''(\rho)$ (see
Eq.~\eqref{eq:eos_chi}). For cold EoS $\kappa=0$. 
Because of the second derivative, in order
not to lose the term $\propto\epsilon''(\rho)$, a quadratic
interpolation must be employed at least. We consider the cubic
interpolation formulas given  either by the standard Lagrangian four
points (centered) stencil or by Hermite polynomials,
e.g.~\cite{Nozawa:1998ak}. In the latter case a table of the
derivative $y'(x)$ must be provided, and it is computed consistently from
Eq.~\eqref{eq:1stlaw}. 
An important point for the accuracy of the interpolation procedure 
is the interpolation of the logarithms. In the tests performed here no
relevant differences are found between Lagrangian and Hermite
interpolations. The method fits also for the cold part of the
hybrid EoS described in Sec.~\ref{sec:eos}.

\subsection{Vacuum treatment}
\label{sbsec:atm}

Of particular importance in the simulation of an isolated object is the
treatment of the vacuum part.
The GRHD equations in vacuum formally do not apply and the
numerical algorithm can not be employed directly because the 
equations to recover the primitives from the conservatives are
singular (see expressions in Appendix~\ref{app:c2p}).

The problem of simulating fluid-vacuum interfaces is quite generic in
fluid dynamics and typically a challenge already
at the Newtonian level~\cite{Toro:1999}. 
A correct, general and robust solution is not currently available 
even without the complication of dynamical spacetimes.
A standard approach, largely employed in the literature, is instead to
replace the vacuum with a minimal atmosphere of density of  
several orders of magnitudes smaller than the typical densities in the 
system. The main claim is that, since the atmosphere density is small,
it will have a negligible dynamical impact. In case of NSs the situation is 
complicated by the presence of gravity and of a stiff fluid.

We implemented a simple vacuum algorithm based on a cold and static
atmosphere. The main ideas come
from~\cite{Font:1998hf,Dimmelmeier:2002bk,Baiotti:2004wn}.
It consists of the following main prescriptions: 
(i)~the atmosphere density value, $\rho_{\rm atm}\equiv
f_{\rm atm}\max\rho$, is chosen as a
fraction, $f_{\rm atm}$, of the maximum density; 
(ii)~the atmosphere pressure and internal energy are chosen according
to the cold part of the EoS of the evolved fluid; 
(iii)~the atmosphere velocity is zero; 
(iv)~the atmosphere is added to initial data in vacuum regions before
starting the time evolution; 
(v)~during the evolution, while recovering the primitive variables, a
point is set to atmosphere if the density is below a threshold 
$\rho_{\rm thr}\equiv f_{\rm thr}\rho_{\rm atm}$. 
Typical values used are $f_{\rm atm}=10^{-10}$ and $f_{\rm thr}=10^{2}$. 
In all our tests we found this approach sufficiently general and robust 
for our purposes.

\subsection{Singularity treatment}
\label{sbsec:sing}

The spacetime singularities treatment in {\bam} is based on the well
known moving puncture method,
e.g.~\cite{Baker:2005vv,Campanelli:2005dd,Hannam:2006vv,vanMeter:2006vi},
which relies on the BSSNOK formulation of the Einstein equations
and on the gauge choice presented in Sec.~\ref{sec:einstein}. 
This method has been proved particularly simple, elegant and robust
and it is widely used in binary black hole
simulations~\cite{Baker:2005vv,Campanelli:2005dd,Brugmann:2008zz} as
well as in matter
simulations~\cite{Duez:2005sf,Shibata:2006bs,Faber:2007dv,Etienne:2007jg,Baiotti:2008ra,Montero:2008yx}. It  
has been shown in fact~\cite{Baiotti:2006wm,Thierfelder:2010dv} that singularities produced by collapsing matter are naturally handled by   
the puncture gauge without particular treatment beyond standard 
artificial dissipation for the metric variables.

As already pointed out in~\cite{Thierfelder:2010dv} however the gauge choice 
alone is not always enough to obtain stable and long term simulations of
collapsing NSs. Unphysical values can be produced by the HRSC scheme
due to numerical errors.  They are typically localized in a
neighborhood of the center of the collapse and appear after the
formation of the apparent horizon. The origin of the problem is clear:
when $v\to1$ (or bigger due to numerical errors) the eigenvalues in 
Eq.~\eqref{eq:eigenv1} become singular and  the formulas to recover the
primitives, too (Appendix~\ref{app:c2p}). 
In order to prevent this our code sets the GRHD eigenvalues to zero if
unphysical values are computed and the numerical fluxes to zero
whenever the velocity becomes larger than the speed of light. We found
this prescription robust in the collapse of both a single star and
of the HMNS produced in binary simulations. 
In~\cite{Thierfelder:2010dv} we tested other possibilities, in particular to set a
ceiling on the Lorentz factor ($W_{\rm ceil}=10^{10}$) if the velocity
becomes larger than the speed of light, as well as hydro-excision. In the case of
single star collapse simulations they lead to comparable results, see
Sec.~\ref{sec:ns:unst}.


\section{NS simulations}
\label{sec:ns}

In this section we validate our code by considering evolutions of single NSs.
Most of the tests presented here were suggested in~\cite{Font:1998hf,Font:2001ew}
and performed later by other groups. They include long term
stability of equilibrium
configurations~\cite{Shibata:2003iy,Baiotti:2010zf,Baiotti:2004wn,Duez:2005sf,Duez:2008rb},   
consistency of linear radial oscillations~\cite{Baiotti:2008nf}, 
migration of an unstable configuration to a stable
one~\cite{Baiotti:2004wn,CorderoCarrion:2008nf}, gravitational
collapse to black-hole~\cite{Shibata:2003iy,Baiotti:2005vi,Montero:2008yx},
and the evolution of a boosted star~\cite{Font:1998hf}. 
In the following we study the performance of different reconstruction
procedures and the convergence of the code evolving stable equilibrium
configurations. We test the EoS interpolation scheme proposed in
Sec.~\ref{sbsec:eos} comparing evolutions with polytropic EoS in
analytic and table form and considering an evolution of a stable
spherical equilibrium model described by a realistic cold EoS.
We demonstrate the ability of the implemented algorithm to handle
shocks (migration test) and the formation of singularities (spherical
collapse).  
The evolution of a boosted star permits the verification of a simple solution of
Einstein equations in a fully dynamical and non-linear case 
and to test the moving boxes technique with the matter scheme.
For this problem we additionally investigate the use of different
values of the $\eta$ parameter in the Gamma-driver shift condition.

\begin{table}
  \caption{Initial data used in single-star evolutions.
    Columns: name, EoS, gravitational (ADM) mass $M$,  
    rest-mass $M_0$, 
    equatorial proper radius $R$, angular momentum $J$ scaled by the
    square of the ADM mass,  
    central rest-mass density $\rho_c$.
   Polytropic models are computed with $\Gamma=2$ and $K=100$.}
  \label{tab:NSid}
  \centering    
  \begin{tabular}{ccccccc}        
    \hline
    Name & EoS  & $M$ & $M_0$ & $R$ & $J/M^2$ & $\rho_c$ [$\times
      10^{-3}$]\\
    \hline
    A0   & polytropic  & 1.400 & 1.506 & 9.586 & 0 & 1.28  \\
    U0   & polytropic  & 1.448 & 1.506 & 5.838 & 0 & 7.993 \\
    F0   & FPS         & 1.400 & 1.566 & 7.367 & 0 & 1.906 \\
   \hline                                   
  \end{tabular}
\end{table}

The main properties of the initial data employed are listed in 
Tab.~\ref{tab:NSid}. Most of them have been previously employed in
such tests. 
Model A0 and U0 are polytropic $\Gamma=2$ spherical
configurations with the same rest mass; A0 is stable while U0
belongs to the unstable branch of the configurations space.
Model F0 is a stable model computed with the FPS EoS~\cite{Friedman:1981,Lorenz1993}.
Spherical configurations are computed with a 2-domain spectral code
described in~\cite{Bernuzzi:2009ex} which employs isotropic
coordinates. 
The values of proper radii in physical units are 
$R=14.156$~km, $8.621$~km and $10.879$~km, respectively, for models
A0, U0 and F0.

The numerical grid set up for these simulations is, except for the
migration and collapse test, such that the finest refinement level 
covers the whole star while the other refinement levels are used
exclusively to the push boundary far away. 
This choice minimizes the effect of the noise due to the interpolation  
between boxes on the matter variables
while it is computational more expensive for a given resolution.
In the case of the migration 
and the collapse test more refinement levels are required to resolve the
expanding or contracting fluid. 
Octant $(x>0,y>0,z>0)$ symmetry is imposed.
We use the 3rd order Runge-Kutta time stepping and a
Courant-Friedrichs-Lewy (CFL) factor of $0.25$. 
In some cases we tested also the use of 4th order Runge-Kutta
finding the same results. 
All the figures in this section and in the following are shown for the
maximum resolution unless explicitly stated otherwise.

\subsection{Stable stars}
\label{sec:ns:stable}

Model A0 and F0 are stable equilibrium configurations, therefore
their evolution is trivial. Since the continuum solution can not
evolve but has to remain in the initial condition, the dynamics of the
numerical solution is governed by truncation errors and by spurious
effects due to the artificial atmosphere. 
This gives the possibility to study long term stability and
convergence of the code. 
Different reconstruction schemes are considered.
The grid is composed of three levels labeled $l=0,..,2$; the
finest box covers entirely the star.  
Simulations are performed employing resolutions of $h_2=0.2,\, 0.3,\,
0.4,\,0.5$ ($h_2=0.295,\, 0.443,\, 0.591,\, 0.738$~km) for the finest
box and last about $10$~ms.  

\begin{figure}[t]
  \begin{center}
    \includegraphics[width=0.45\textwidth]{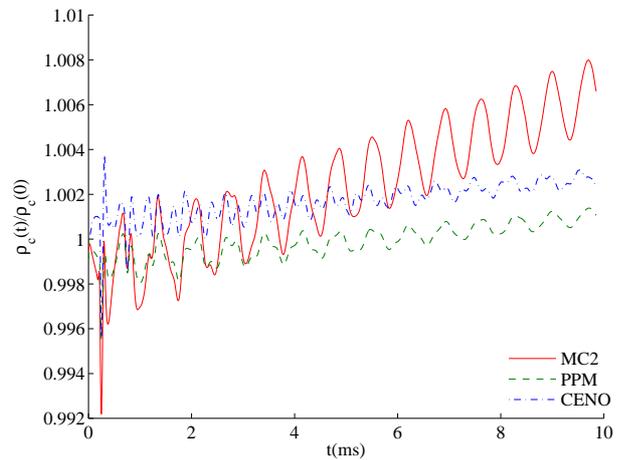}
    \caption{\label{fig:stable:rhoc} Evolution of the central rest-mass
      density of model A0. The picture shows the normalized value
      $\rho_c(t)/\rho_c(0)$ for different reconstruction schemes.}
  \end{center}
\end{figure}

\paragraph*{Long term stability and convergence.}
We discuss here the simulations of model A0. 
It is well known, e.g.~\cite{Font:1998hf}, that numerical errors trigger
small amplitude pulsations of the star which oscillates at proper mode
frequencies. The phenomenon is depicted in Fig.~\ref{fig:stable:rhoc}
where the evolution of the central rest mass density normalized to its
initial value is shown for different reconstruction schemes.
The figure demonstrates also the ability of the code to maintain the
initial configuration. For instance the pulsations amplitude is less
than 0.5\% over $10$~ms. Two frequencies dominates the pulsations: $\nu_{\rm
  F}=1421\,$~Hz and $\nu_{\rm H}=3959\,$~Hz. They agree within
the errors estimated from the output time sampling ($2$\%) with the
fundamental radial linear mode and its first overtone as computed by 
perturbative methods~\cite{Kokkotas:2000up,Bernuzzi:2008fu}.   
The figure highlights a secular drift in the case that MC2 reconstruction
is adopted. A similar drift is observed also in simulations with PPM and
CENO at lower resolutions.  
This secular drift is a feature related to the evolution of
the geometry together with the fluid since, if we perform simulations
in the Cowling approximation (metric variables not evolved), it is 
almost absent at all the resolution. 
The amplitude of the pulsations is also larger in the case of MC2
indicating that the truncation errors are bigger as expected for a
linear reconstruction. Notice that in the simulations with CENO
reconstruction the overtone of the radial mode at frequency $\nu_{\rm H}$ 
is more clearly visible.
  
During preliminary tests with CENO reconstruction we observed a loss
of stability between 2 and 6~ms depending on the resolution employed.
We found that for a compact star evolution and a standard
implementation of the CENO reconstruction, the limiter tends to select 
in some points the lower order sub-stencils.
A similar effect is also discussed in~\cite{Anderson:2007kz}. 
The problem is easily fixed in our set up by choosing a different
weight in the weighted differences between the linear reconstruction and
the quadratic polynomial with centered stencil. Specifically we set
the free parameter $\alpha^0=0.7$ in Eq.~A.6 of~\cite{Zanna:2002qr} to
$\alpha^0=0.1$. In this way the limiter selects the central higher order 
stencil. We found this solution sufficient for all the problems and at 
all the resolutions considered in this work. We note
that also the PPM reconstruction has several free parameters to tune.
We did not attempt tuning but instead we use the prescription 
given in~\cite{Marti:1996}. 

\begin{figure}[t]
  \begin{center}
    \includegraphics[width=0.45\textwidth]{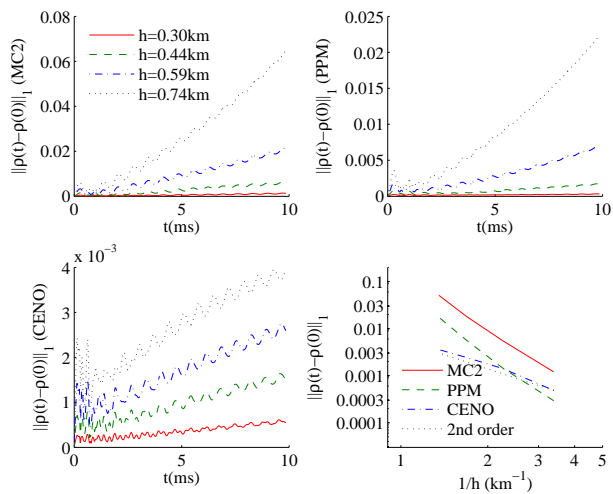}
    \caption{ 
      Convergence in L1 norm of evolutions of model A0. 
      The first three panels from the top-left to the bottom-right
      show the evolution of the L1 distance
      $||\rho(t)-\rho(0)||_1$.
      Different lines refer to different resolutions.
      Different panels refer to the
      three different reconstructions considered: MC2, PPM and
      CENO. The last panel (bottom-right) shows in a log-log plot the 
      L1 distances as a function of $1/h$ at $t=8$~ms for the three
      reconstructions MC2 (solid-red line), PPM 
      (dashed-blue line) and CENO (solid-green line).
    }
    \label{fig:stable:L1rho} 
  \end{center}
\end{figure}

Next we discuss convergence. 
Fig.~\ref{fig:stable:L1rho} reports the evolution of the L1 distance
between the rest-mass density evolved and the initial data. 
The different curves refer to different resolutions; at every time the
difference between two curves behaves as $||\rho(t)-\rho(0)
||_1\propto h^{r}+\mathcal{O}(h^{r+1})$ since the 
initial data represent also the solution for the evolution problem.
The three panels from top left to bottom left refer to different
reconstructions. In all the cases we observe convergence with
increasing resolution; 2nd order convergence is found at early times
$t\sim2$~ms for all the methods. 
At late times however the MC2 and PPM
reconstructions show larger truncation errors and the curves present a
quadratic behavior. This leads to apparent over-convergent results, 
which indicates the simulations are not yet in the convergent regime.  
By contrast the use of CENO reconstruction gives 2nd order convergence
over the whole simulated time.   
The last panel (bottom right) summarizes the observed convergence rate
by showing in a log-log plot the L1 distances as a function of $1/h_2$ at 
$t=8$~ms and for different reconstructions.
The slope of the lines gives the convergence rate. The over-convergent 
behavior for MC2 and PPM as well as the 2nd order convergence of CENO
are evident. 

\begin{figure}[t]
  \begin{center}
    \includegraphics[width=0.45\textwidth]{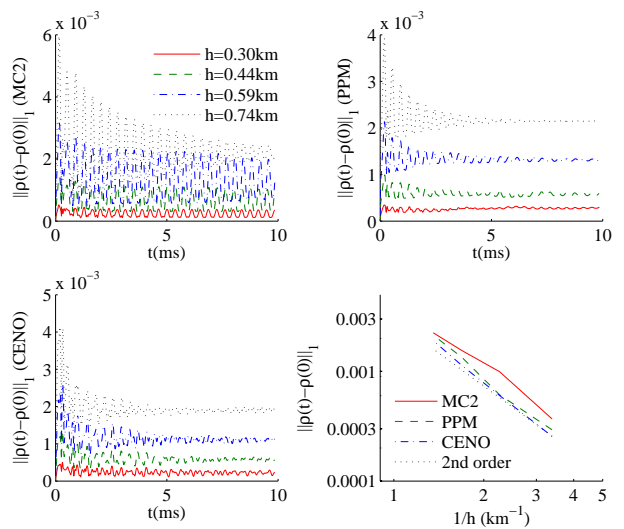}
    \caption{  Same as in Fig.~\ref{fig:stable:L1rho} 
      but results are computed in the Cowling approximation.}
    \label{fig:stable:L1rho_COW}
 \end{center}
\end{figure}

To interpret these results we recall here that in a HRSC scheme the
truncation errors strongly depend on the degree of smoothness of the
solution and on the specific limiter employed.  
While the formal convergence rate of the methods employing different
reconstruction is the same (2nd order), the truncation errors for this problem are
simply different in the three cases. PPM and MC2 reconstructions are
only first order at smooth extrema thus they are expected to be less
accurate on smooth solution than CENO, but more robust in the presence of
strong shocks. More importantly the apparent over-convergence
behavior is not simply related to the HRSC method employed to solve 
the GRHD equations, but it seems a genuine aspect of evolving the 
GRHD equations coupled to the Einstein equations.  
To show this we consider Fig.~\ref{fig:stable:L1rho_COW}, which is the
same as Fig.~\ref{fig:stable:L1rho} but in the Cowling approximation.  
From an inspection of the figure it is clear that, once the metric is not
evolved, all the reconstructions perform quite similarly   
showing perfect 2nd order convergence in L1 norm at all times. 
Specifically one finds that CENO and PPM have similar
performance and that for all the reconstructions the convergent regime is
reached at the considered resolutions. We conclude that the slower
convergence observed for MC2 and PPM in the case the metric is evolved
is due to a combination of numerical errors from various part of  
the algorithm rather than to an effect related only to the
reconstruction methods. 

\begin{figure}[t]
  \begin{center}
    \includegraphics[width=0.45\textwidth]{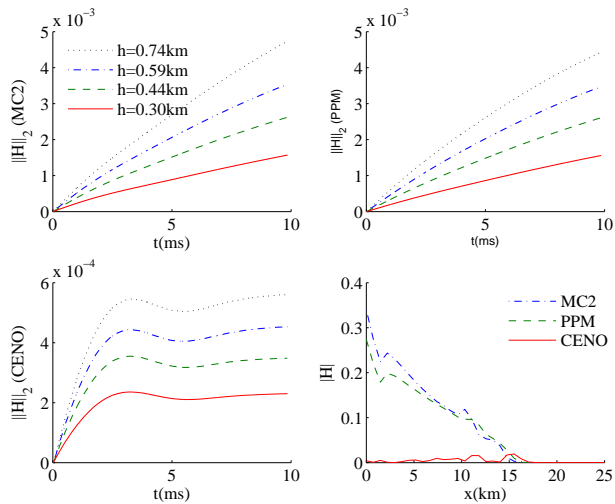}
    \caption{
      Hamiltonian violation during the evolution of model A0. 
      The first three panels from the top-left to the bottom-right
      show the time evolution of the 2-norm of the Hamiltonian 
      constraint computed on the finest grid level which fully contain
      the star. Different lines refer to different resolutions.
      Different panels refer to the
      three different reconstructions considered: MC2, PPM and
      CENO. The last panel shows the profile in the $x$ direction of
      the Hamiltonian constraint at time $t\sim10$~ms for the maximum
      resolution and three reconstructions MC2 (solid-red line), PPM
      (dashed-blue line) and CENO (solid-green line). 
      Note the different scales of the plots.
   }
   \label{fig:stable:ham} 
  \end{center}
\end{figure}

We finally comment on rest-mass conservation. In all the
simulations reported here we observe that the deviation runs from a maximum 
of $\Delta M_0/M_0\sim 10^{-2}$ with MC2 reconstruction at the lowest resolution 
to a minimum of $\Delta M_0/M_0\sim 10^{-6}$ for CENO at the highest resolution.
This remarkable result is a well known consequence of
the use flux-conservative methods.

\paragraph*{Constraint violation.} 
The Einstein constraints are violated
at the level of the numerical error independently of the formulation
or the method adopted. When free-evolution schemes are adopted,  
the constraints are only monitored (not solved) and typically the
violation (i)~grows in time, (ii)~converges with increasing
resolution.  
In our simulations we observe this behavior. The biggest violation on 
the grid are found in the region covered by the matter  
and, at least at the initial time, at the boundary. 

Fig.~\ref{fig:stable:ham} summarizes our findings.
The three panels from top-left to bottom-left show the evolution of
the L2 norm of the Hamiltonian constraint for several resolutions in the cases 
when MC2, PPM and CENO reconstruction are employed. The violation 
converges to zero in all the cases. 
From the figures is clear that the absolute value of the violation is 
larger for the MC2 reconstruction while it is about a factor 10
smaller for CENO reconstruction when compared to MC2 and PPM. 
A summation of effects and errors as described in the previous
paragraph contributes to this behavior. 
It is difficult to clearly identify them due to the complexity of the
equations and of the numerical method employed.

We observe that, if mesh refinement is not employed but only  
one box corresponding to the 3rd level is used, then the norm of the  
Hamiltonian constraint shows an anti-convergence behavior at early 
times. This feature is due to the initial non-convergent constraint 
violation at the boundary, related to the use of Sommerfeld conditions~\cite{Ruiz:2010qj}, 
which dominates the violation in the interior 
at early times. The use of mesh refinement is thus important to 
minimize the effect of the boundary conditions.

The bottom-right panel displays the spatial profile in the $x$ direction
of the Hamiltonian constraint at late time, $t\sim10$~ms, for
simulations employing the highest resolution and different
reconstruction schemes. The Hamiltonian violation accumulates in time 
in the region of the matter. Once again the difference in the results is clear 
between the CENO reconstruction and PPM and MC2.
In case of the unigrid simulations mentioned above the bulk violation 
dominates over the boundary violation after the first ms of evolution 
in the case that PPM or MC2 reconstruction is employed. 

These tests convinced us to use the CENO reconstruction (see also
Sec.~\ref{sbsec:bns:dyn}). We stress that if a different setting  
is used, e.g.\ a different mesh, method to solve Einstein equations,
numerical flux, etc. or simply much higher resolutions, the picture may 
change.  

The constraint accumulation and boundary condition effects discussed
here are related to the use of BSSNOK and the Sommerfeld boundary condition.  
They have been studied in detail in~\cite{Bernuzzi:2009ex,Ruiz:2010qj}
where the Z4c formulation was proposed as an alternative. 
Ref.~\cite{Bernuzzi:2009ex,Ruiz:2010qj} are restricted to 
spherical symmetry, but we will investigate in the future the use of Z4c 
in 3D.

\paragraph*{Interpolation of EoS tables.}
In order to establish the performance of the interpolation scheme for  
EoS 1D tables described in Sec.~\ref{sbsec:eos}, we consider 
simulations of model A0 and of model F0 employing tables. 

The evolutions of model A0 performed both with the analytic expression
for the polytropic EoS and with a test-table do not significantly
differ. As an example the central rest mass density show differences
smaller than $0.01$ \% over $10$~ms of simulated time for a resolution
of $h_2=0.2$ ($h_2=0.295$~km). Performance is clearly affected by the
use of tables:  we observe on average a slow down of about 10 \% but
always less than 20\%. The use of tables with different entries,
e.g.~126 and~1024, does not lead to differences in these simulations. 

In order to obtain an accurate evolution of model F0 we employ a table 
generated from the analytic fits of the FPS EoS~\cite{Haensel:2004nu,Shibata:2005ss}. 
Initial tests with publicly available tables showed a progressive
accumulation  of round off errors as simulation time advances that
eventually led to a failure during the recovery of primitives.   
The reason is the low accuracy of the table entries (in some cases few
digits). Once tables from the fits are employed, the simulations are
stable and accurate as in the polytropic case. For what concerns the proper
radial frequencies of F0, we estimate $\nu_{\rm F}=3045$~Hz and
$\nu_{\rm H}=7232$~Hz, with a resolution of $h_2=0.2$ and a total
simulated time of $7$~ms. They agree within few percent with
perturbative results computed in~\cite{Bernuzzi:2008fu}. As expected
in this case the evolutions with the cold table and with the
corresponding hybrid EoS do not differ significantly, while we observe
some small differences due to the different atmosphere treatment.

\subsection{Unstable stars}
\label{sec:ns:unst}

\begin{figure}[t]
  \begin{center}
    \includegraphics[width=0.45\textwidth]{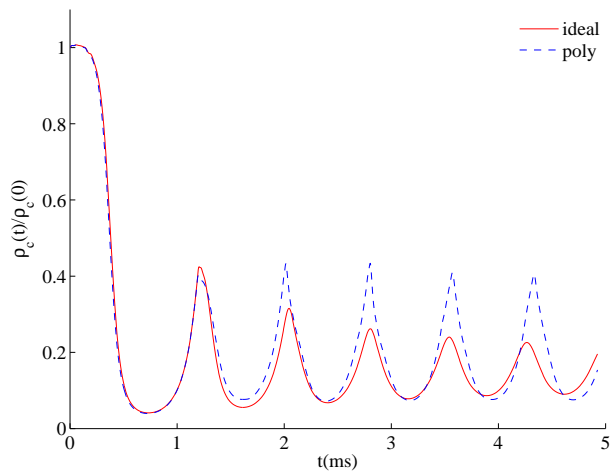}
    \caption{ Evolution of the central rest-mass   
      density in the migration test. The picure shows the normalized value
      $\rho_c(t)/\rho_c(0)$ in the case of an ideal gas EoS (red solid line)
      and polytropic EoS (blue dashed line).}
      \label{fig:migr:rhoc}
  \end{center}
\end{figure}

We discuss the evolution of model U0. Since it is an unstable
configuration a small perturbation can cause either the migration
towards a stable model of the same rest mass or the collapse to
a black-hole. In practice truncation errors are enough to trigger the
migration while the further addition of a small radial perturbation
(bigger than the truncation errors) leads to the collapse.
In the following we will consider both cases.
The simulations presented here were performed with CENO
reconstruction. A direct quantitative comparison of the results for
the collapse with an independent 1D spherical code can be found
in~\cite{Thierfelder:2010dv}. 

\begin{figure}[t]
  \begin{center}
    \includegraphics[width=0.45\textwidth]{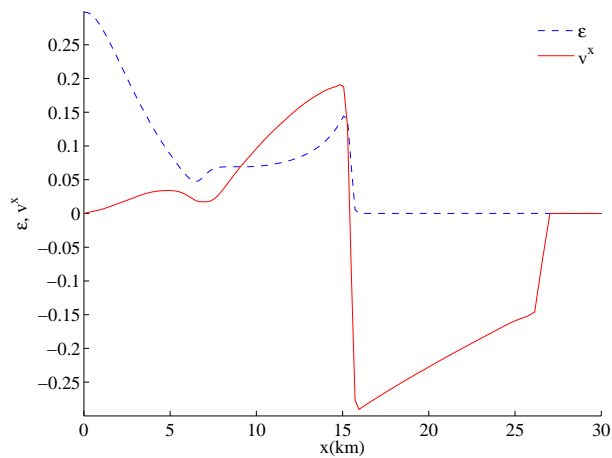}
    \caption{Shock formation in the migration
      test. The picture shows the profiles of the specific internal
      energy $\epsilon$ and of the component $v^y$ of the velocity at 
      time $t\sim1.1$~ms, i.e.\ soon after the shock formation. 
      Note that only $1/10$ of the numerical grid is shown.} 
      \label{fig:migr:shock} 
  \end{center}
\end{figure}

\paragraph*{Migration.} The dynamics of a migrating star is described
in detail in~\cite{Font:2001ew,CorderoCarrion:2008nf}. Within the
first $0.5$~ms the star rapidly expands while the central density
decreases by a factor 8 to $\rho_c\sim1.4\times10^{-3}$; a phase of
strongly nonlinear pulsations around the stable configuration then
starts. 
The configuration reached is in fact the perturbed model A0, where the
difference in the ADM mass between A0 and U0 has been converted into
the kinetic energy of the pulsations. If the polytropic EoS is
adopted, thus enforcing an isentropic evolution, oscillations can be
damped by the effective numerical viscosity of the HRSC scheme and by
spurious interaction with the atmosphere. If the ideal gas EoS is
employed, thus allowing shock heating, a shock forms during the first
pulsation at the interface between the inner core and the infalling
mantle (outer lower density matter). The inner core then bounces and
expands again several times feeding the shock with kinetic energy, which
dissipates it into thermal energy. As a result the oscillation
amplitudes decrease.  
Fig.~\ref{fig:migr:rhoc} shows this behavior captured by our
simulation; the evolution of central rest-mass density is plotted for
the two different EoS. The amplitude of the pulses remains
approximately constant during the simulation in the case of
polytropic EoS, indicating the spurious numerical effects are
negligible, while it decreases as expected in the case of the ideal gas
EoS. The capture of the shock dynamics is demonstrated in
Fig.~\ref{fig:migr:shock} which shows the profile in the $x$ direction of
the specific internal energy and of the $x$ component of the velocity
at the time $t\sim1.1$~ms, i.e.\ just after of the shock formation when
the shock wave is moving outward. The figure clearly indicates how part of
the matter is expanding outwards while the low density mantle is
infalling towards the symmetry center. Note the steep profile of the
velocity. We observe that, in the ideal gas EoS 
case, the expansion of the star extends to the boundary of the
numerical grid and a fraction of the mass, 1-2~\% percent over the
simulated time, is lost outside~\footnote{
  For this reason the expected end state of the simulation is not
  exactly model A0 but a model with a smaller mass.}.   
The grid covers a cube of side $200$ ($295$~km) and five 
levels are employed with a maximum resolution of 
$h_4=0.15$ ($h_4=0.221$~km) and a minimum of $h_0=2.4$ ($h_0=3.544$~km). 
The region within a coordinate
radius of about $r\sim50$ ($74$~km) is entirely resolved by the two
finest refinement levels. 
By contrast, in the isentropic case, the expansion reaches a maximum
coordinate radius of $r\sim25$ ($37$~km) and the rest mass is
conserved as in the stable star tests. 

\begin{figure}[t]
  \begin{center}
    \includegraphics[width=0.45\textwidth]{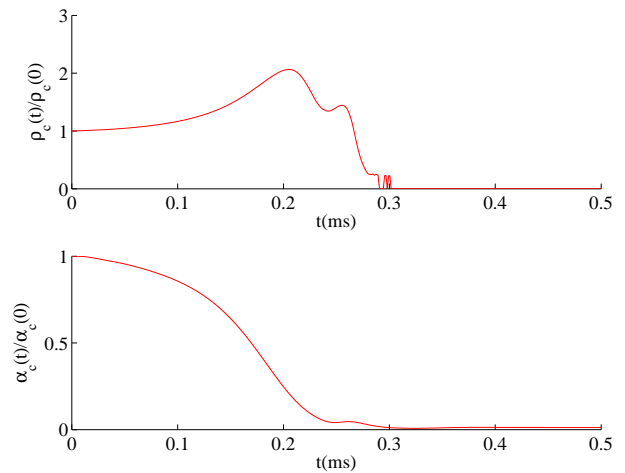}
    \caption{Dynamics of the collapse
      to a black hole with puncture gauge. The picture shows the evolution
      of the central rest-mass density (top panel) and of the central
      lapse (bottom panel). Both are normalized by the central value
      of the initial data.
    }
    \label{fig:coll:dynamics} 
  \end{center}
\end{figure}

\paragraph*{Collapse.} The collapse is triggered by introducing a
radial perturbation of the velocity field with an amplitude larger
than truncation errors. Since we are only interested in testing the
ability of the code to handle the formation of singularities, to
keep the set up simple we do not solve the
constraints after imposing the perturbation. In our experience this
procedure does not introduce relevant unphysical effects, while it is
clearly an inconsistent way to solve Einstein equations (see also
discussions in~\cite{Baiotti:2004wn,Baiotti:2007np,Anderson:2007kz}). 
Furthermore the results obtained are in close qualitative agreement
with~\cite{Font:2001ew,CorderoCarrion:2008nf} where the
constraints were solved.  
For these simulations we employ an eight level grid with maximum
resolutions of $h_7=0.05,\, 0.03125,\, 0.025$ ($h_7=0.0738,\, 0.0461,\,
0.0369$~km) which cover a cube of side $1.164$ ($1.72$~km).

The collapse happens in the first $0.3$~ms of simulation: the matter
falls towards the symmetry center while the metric varies rapidly
adapting itself to the matter distribution. At about 
$t_{\rm AH}\sim 0.23$~ms an apparent horizon forms with an
initial coordinate radius of $r_{\rm AH}(t_{\rm AH})\lesssim 0.96$
($1.41$~km or $0.66$~M). Part of the matter is
outside of it and then rapidly accreted. 
Fig.~\ref{fig:coll:dynamics} shows the evolution of the central
rest-mass density (top panel) and that of the central lapse (bottom
panel). The central density increases of a factor two at $t_{\rm AH}$
while the lapse collapses towards zero. After $t_{\rm AH}$ the gauge
conditions in Eq.~\eqref{eq:lapse_1log} and Eq.~\eqref{eq:beta_Gdriv}, 
in particular the Gamma-driver for the shift,
play the main role in handling the singularity. As described in detail
in~\cite{Thierfelder:2010dv}, the shift condition deforms the spatial 
coordinates pushing the proper (Schwarzschild) radii 
$r_{\rm Schw}\lesssim 1.7$ ($2.55$~km or 1.2~M) 
outside the innermost point of the numerical grid (see also the top panel 
of Fig.~6 in~\cite{Thierfelder:2010dv}). Consequently: 
(i)~matter effectively disappears from the numerical grid; 
(ii)~the end state of the collapse is the trumpet slice of
Schwarzschild~\cite{Brandt:1997tf,Hannam:2006vv,vanMeter:2006vi,Brown:2007tb}.
The central rest-mass density in Fig.~\ref{fig:coll:dynamics}
reaches the atmosphere value about $\Delta t\sim0.1$~ms after $t_{\rm
  AH}$. Some spurious effects are visible during this time interval
and they are related to the numerical treatment of the matter fields
discussed in Sec.~\ref{sbsec:sing}. We experimented with 
the alternative treatments described in Sec.~\ref{sbsec:sing}, finding
unimportant quantitative differences only in the short phase before all
the matter disappears except for the atmosphere.

\begin{figure}[t]
  \begin{center}
    \includegraphics[width=0.45\textwidth]{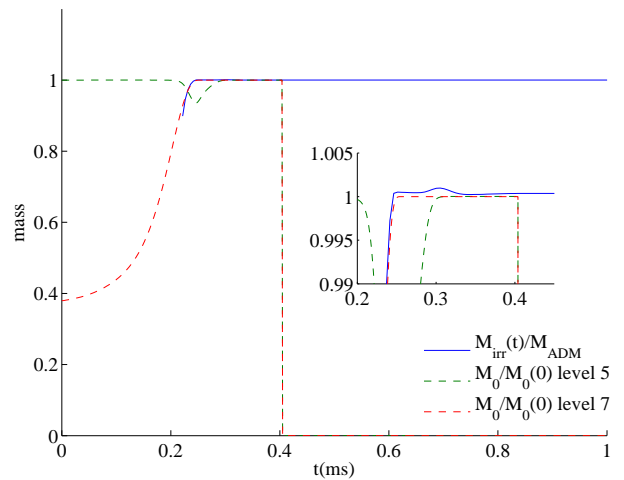}
    \caption{Evolution of different mass quantities in the collapse 
      test. The picture shows the irreducible mass
      (solid line) of the final black hole and the rest mass of the
      matter (dashed line) normalized to their initial values at level 5 and 7.
    }
    \label{fig:coll:mass}
  \end{center}
\end{figure}

The behavior of the matter is also demonstrated in
Fig.~\ref{fig:coll:mass} which shows the evolution of the 
irreducible mass of the black hole normalized to the ADM mass of the
system (solid line) and of the rest mass normalized to the initial
value (dashed lines). The irreducible mass of the black hole is
obviously zero at the beginning and after $t_{\rm AH}$ it rapidly
reaches a value corresponding to the initial ADM mass of the system. 
The rest mass is shown computed on the finest
refinement level (level 7) for the best resolution ($h_7=0.025$) and on 
a coarser one (level 5) with a resolution of 
$h_5/4$. Since the matter is contracting, the mass on the level 7 
initially increases and after $t_{\rm AH}$ the level contains all the
matter. Level 5 instead always contains all the matter but after $t_{\rm AH}$
does not resolve it properly.
During the collapse the rest-mass is conserved up to $0.05\%$ 
while, after $t_{\rm AH}$, it rapidly decreases to the atmosphere value, 
$M_0(t>t_{\rm AH})\sim10^{-6} M_0(0)$.  

\subsection{Boosted star}
\label{sec:ns:boost}

We discuss here the evolution of model A0 boosted in the $x$
direction at a speed of $v=0.5$ corresponding to a
Lorentz factor of $W=1.16$. The setup of the initial data is as
described in~\cite{Font:1998hf}.  
The test is interesting because it gives the possibility to experiment
in a simple scenario with several points:  
fully dynamical and nonlinear evolutions, the performance of the
moving boxes with matter, and different gauge conditions. 
We use five refinement levels, the moving one is the finest 
which entirely covers the star. The different resolutions employed for
the finest box are $h_4=0.4,\, 0.278,\, 0.208$ ($h_4=0.588,\, 0.408,\,
0.306$~km). 

\begin{figure}[t]
  \begin{center}
    \includegraphics[width=0.45\textwidth]{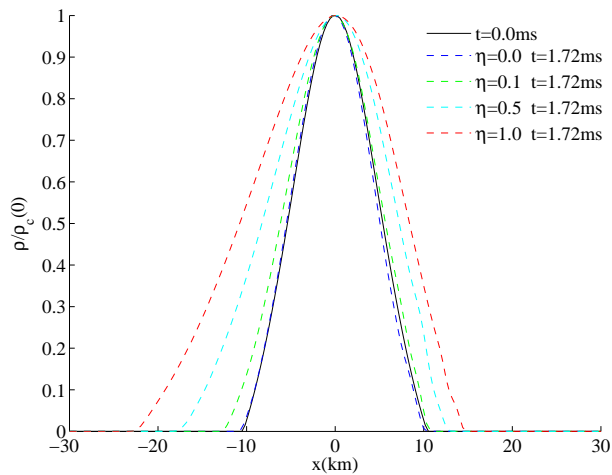}
    \caption{Profile of rest-mass density of
      the evolved boosted model A0. The picture shows the rest-mass density
      profile normalized to the initial value at time $t=1.72$~ms for
      evolutions corresponding to different values of the $\eta$
      parameter in the Gamma-driver shift (dashed colored lines). 
      The profiles are shifted back to the initial position of the
      star, the initial data is also plotted (black solid line).}
    \label{fig:boost:rhoeta} 
  \end{center}
\end{figure}

The solution of the evolution problem depends on the gauge conditions employed. 
If we choose to simply advect lapse and shift,  
the solution is analytic and it is just a time shift of the initial
metric and matter profiles. If the conditions in
Eq.~\eqref{eq:lapse_1log} and Eq.~\eqref{eq:beta_Gdriv} 
are used, the solution is not analytic. 
In order to investigate the numerical solution obtained under the
1+log and Gamma-driver condition we consider evolutions with different
values of the parameter $\eta$.  
A similar investigation has been carried out for BNS and it is
presented in Sec.~\ref{sbsec:bns:dyn}. Fig.~\ref{fig:boost:rhoeta} summarises  
our findings. It shows the profiles of the rest-mass density in the $x$
direction at the final time of $t=1.72$~ms for evolutions using
different values of $\eta$. All the profiles are shifted back to the
initial position of the star, the initial profile is also plotted. As
apparent from the figure, 
the choice $\eta=0$ corresponds to the case closest to the
analytic solution, while for higher values the star profile is
progressively more distorted in the direction of motion. 

\begin{figure}[t]
  \begin{center}
    \includegraphics[width=0.45\textwidth]{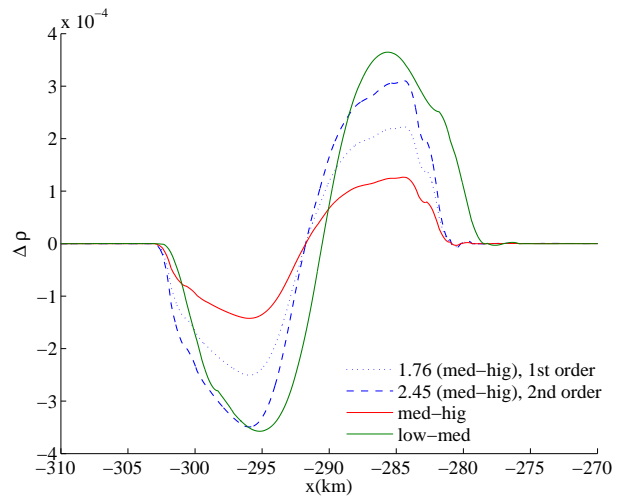}
    \caption{Self-convergence test of the evolution
      of a boosted star. The pictures shows the differences between the
      profiles of the rest-mass density at $t=3.95$~ms evolved with
      different resolutions. The labels of the solid lines ``low'',
      ``med'' and ``hig'' in the legend correspond to resolutions
      $h_4=0.4,\, 0.201,\, 0.278$.       
      The difference between medium and high resolution is scaled for
      2nd order (dashed line) and 1st order (dotted line). 
      The gamma-driver condition employs $\eta=0$.
    }
    \label{fig:boost:conv} 
  \end{center}
\end{figure}

We finally comment about convergence. 
Fig.~\ref{fig:boost:conv} shows the point-wise self-convergence the
spatial profile of $\rho$ at $t=3.95$~ms for $\eta=0$. 
We show the differences between the rest-mass density profile computed
at different resolutions. The difference between the medium, 
$h_4=0.278$, and the high resolution, $h_4=0.208$, 
is  scaled by a factor corresponding to 1st (dotted
line) and 2nd (dashed line) order convergence. Point-wise convergence is
lost at early times but the magnitude of the errors scales at about
2nd order thus indicating convergence in the L1 norm as expected. 
Note that the evolution time presented here is a factor of about
$10^{4}$ longer than simulations in~\cite{Font:1998hf}.


\section{BNS simulations}
\label{sec:bns}

In this section we discuss our BNS simulations.
We focus on a simple equal-mass irrotational configuration
evolved with polytropic and ideal gas EoS. In both evolutions the
formation of a HMNS is observed, but in the isentropic case it
collapses after about $3$~ms producing a Kerr BH surrounded by a disk
of mass $M_{\rm d}\lesssim 2\times10^{-2}M_0$, while in the other case
the HMNS survives for about $9$~ms before collapsing.
We give an overview of the dynamics and discuss the impact of using 
different resolutions and different reconstructions schemes. 
We present results concerning the use of different values
of the damping parameter $\eta$ in the shift condition: while small
values of $\eta$ lead to less coordinate eccentricity during the inspiral, 
they reduce the coordinate size of the final BH. 
We present and characterize the GWs computed from the simulations.
We compute the actual GWs degree of  freedom (metric waveforms) by
means of the post-processing algorithms described
in~\cite{Damour:2008te,Baiotti:2008nf} and~\cite{Reisswig:2010di} and
compare their performance.
Convergence tests are performed indicating 2nd order convergence of
the inspiral waveforms without any time-shifting procedure. 
For the first  time in BNS simulation, we estimated precise error-bars
on the waveforms by extrapolating the results in resolution.

\subsection{Initial data and numerical settings}
\label{sbsec:bns:setup}

\begin{table}
\caption{Parameters of the initial binary
  configuration. Columns: name, ADM mass $M$, rest mass $M_b$ and ADM
  mass $M_\star$ of each star in isolation, angular momentum $J$
  scaled by the square of the ADM mass, gravitational wave frequency
  $\omega_0$, proper separation $d$, central density of each
  star $\rho_c$. The parameters in the polytropic EoS are $\Gamma=2$
  and $K=123.6489$.}  
  \label{tab:idbns}
\centering    
\begin{tabular}{cccccccc}
  \hline
  name  & M & $M_{\rm b}$ & $M_\star$ & $J/M^2$ &$\omega_0$ [Hz] & d
  & $\rho_c$ [$\times10^{-4}$]\\
  \hline
  G2P14 & 2.998 & 1.625 & 1.499 & 0.4450 & 589 & 36.582 & 9.569 \\
  \hline
\end{tabular}
\end{table}

We employ as initial data quasi-equilibrium configurations of 
irrotational equal-mass binaries in quasi-circular orbits. 
These initial configurations are computed with a multi-domain spectral 
code which solves the Einstein constraint equations under the
assumption of a conformally flat metric. The code is based on the 
LORENE library~\cite{LORENE} and provided by the
NR group in LUTH (Meudon). These initial 
data represent to date the most accurate computation of equilibrium
BNSs and they are publicly available on the web.
The assumption of irrotational flow is believed to be astrophysically
realistic since the spin period of the neutron star is larger then
the orbital period in the late inspiral. In other terms, the time of
coalescence due to gravitational radiation is shorter than that of
synchronization due to viscosity. Because the late inspiral and merger
phases are expected to happen a very long time after the birth
of the binary system ($\sim10^8$~yr), the temperature of the matter
can be assumed to be negligible. The neutron stars in the equilibrium
system are thus described by the cold EoS. 

The properties of the initial configuration that we consider in the
following are summarized Tab.~\ref{tab:idbns}. The binary has ADM
mass $M=2.998$, angular momentum $J=8.836$ and proper separation
$d\simeq36.582$~($54$~km), thus the compactness of the system is
$M/d\simeq0.08$. The coordinate separation is $30.457$~($45$~km). 
The rest-mass and ADM mass of each star in isolation ($d\to\infty$,
spherical configuration with same rest-mass) are
$M_{\rm b}=1.625$ and $M_\star=1.499$, respectively. Note the
notation for the rest-mass of the star in isolation, $M_{\rm b}$, and for
the rest-mass of the binary, $M_0$. The compactness of each star in
isolation is $M_\star/R=0.14$. The equilibrium configuration was first
computed  in~\cite{Gourgoulhon:2000nn,Taniguchi:2002ns}.

As a test for the implementation we performed evolutions with other initial
configurations. In particular we considered an irrotational
equal-mass binary described by the APR EoS~\cite{Akmal:1998cf}
computed in~\cite{Gourgoulhon:2000nn,Bejger:2004zx} and evolved
in~\cite{Shibata:2006nm,Kiuchi:2009jt}. The evolutions were performed
using both an APR table based on the analytic fit
of~\cite{Shibata:2006nm} and its hybridization with a $\Gamma=2$ ideal
gas EoS. The robustness of the procedure described in Sec.~\ref{sbsec:eos} 
has been checked also in the case of BNSs. All these
tests were performed at grid configuration L, M and H0 (see below),
thus we do not discuss them in the following. 
Another configuration whose evolution was explored is an irrotational
equal-mass binary described by a polytropic EoS and with ADM mass $M=3.251$ 
and compactness of each star $M_\star/R=0.16$~\cite{Gourgoulhon:2000nn,Taniguchi:2002ns}.
As expected evolutions with both  polytropic and ideal gas EoSs led to
a prompt collapse to a BH and the main evolution happens in
$11$~ms. Results are comparable to the findings in~\cite{Baiotti:2009gk}.  
Since the evolution of model G2P14 has a more complicate post-merger dynamics 
we focus on the results obtained from these evolutions. 
They evidently represent a more stringent test for the code.

\begin{table}
\caption{Grid configurations used in BNS
  simulations. 
  Columns: identifier, number of grid levels (boxes), 
  number of moving boxes, resolution of finest box (dimensionless and
  km), number of points in finest box, resolution of coarsest box
  (dimensionless and km), number of points in coarsest box.}
  \label{tab:gridbns}
\centering    
\begin{tabular}{cccccccccc}
  \hline              
  name  & RL & MRL & $h_5$ & $h_5$ [km] & $N_5$ &
  $h_0$ & $h_0$ [km] & $N_0$ & boundary\\
  \hline
  L    & 6 & 4 & 0.500 & 0.74 & 40  & 16.0 & 23.6 & 80  & 945\\
  M    & 6 & 4 & 0.400 & 0.59 & 50  & 12.8 & 18.9 & 100 & 945\\
  H0    & 6 & 4 & 0.313 & 0.46 & 64  & 10.0 & 14.8 & 128 & 945\\
  H1    & 6 & 4 & 0.250 & 0.37 & 80  & 8.0  & 11.8 & 120 & 709\\
  H2    & 6 & 4 & 0.200 & 0.29 & 100 & 6.4  & 9.4  & 210 & 982\\
  H3    & 6 & 4 & 0.156 & 0.23 & 128 & 5.0  & 7.4  & 260 & 960\\
\hline                                   
\end{tabular}
\end{table}

In Tab.~\ref{tab:gridbns} we report the grid configurations used for 
the evolution simulations. 
Simulations with configurations L and M can be run on a small
machines. They need between 8 and 16 processors with 1GB of memory per
core. While they can be carried out without any problem,
thus proving the flexibility of the code, the results are too
inaccurate to be considered for a sensible analysis. 
Convergence can be measured only in simulations employing
configurations H0 as shown in the following.  
For each grid configuration the finest refinement level covers 
each star entirely. The latter is resolved with $128$ points
in runs H3 and $64$ points in H0. Let us briefly comment about the grid setting 
used here and those used for BBH simulations, e.g.~\cite{Brugmann:2008zz,Husa:2007hp,Hannam:2007ik}.
For BBH simulations roughly half the number of points per direction than here is used 
together with more grid levels (typically 9-11 levels). Higher resolutions are 
reached in the finest grid level in order to resolve the punctures, 
while comparable (or less) resolutions are used on level 5. Therefore
the horizon of the final BH in BBH simulations is resolved typically 
on level 6 or 7 with a resolution about two to four times better than in 
BNS simulations. One important consequence is that the precision 
of the apparent horizon finder is affected. 

The performance of the code for each grid configuration is reported
in Tab.~\ref{tab:perf}. The BNS runs described in this paper require an 
average speed of $\sim3$~M/hr (H3) on 128 processors in the LRZ cluster
(Munich) and $\sim9$~M/hr (H0) on 32 processors. In physical units
$10$~M of the configuration selected corresponds to $\sim0.05$~ms of
simulation. 
We performed scaling tests up to 512 processors finding good scaling 
properties by using larger grid setups and higher resolutions.
Production runs with resolutions of $h_5\sim0.12-0.10$ on
256-512 processors are thus definitely feasible in reasonable times
with our code. 
Here we did not run such kind of simulations because of our computer time
restrictions. 

Our grid settings are similar to those of other
codes~\cite{Baiotti:2010ka,Giacomazzo:2010bx}. 
The highest resolution employed here is
$30$~\% lower than the maximum resolution used to date on BNS
simulations employing mesh-refined-Cartesian-grid-based 
codes~\cite{Baiotti:2010xh,Giacomazzo:2010bx} and comparable to~\cite{Kiuchi:2009jt}. 

If not explicitly stated the data discussed refer to simulations
employing RK3 time stepping, CFL factor of 0.25 and CENO reconstruction.

\begin{table}[t]
  \caption{ Performances of BNS runs.
    Columns: grid configuration, number of processor, cpu memory, total
    runtime, average speed. Runs last to $t=5000$($1666~M$) 
     on LRZ Munich. The numbers include LORENE initial data 
     interpolation and checkpointing.} 
  \label{tab:perf}
  \begin{tabular}{ccccc}
    \hline              
    Grid  & nproc & mem [Gb] & time [CPU hr]& speed [M/hr]\\
    \hline 
    H0 & 32  &  90 & 192 & 8.7 \\
    H1 & 32  &  96 & 268 & 6.2 \\
    H2 & 128 & 120 & 254 & 6.5 \\
    H3 & 128 & 165 & 480 & 3.5 \\
  \hline
  \end{tabular}
\end{table}

\subsection{Overview of  the dynamics}
\label{sbsec:bns:dyn}

\begin{figure*}[t]
  \begin{center}
\includegraphics[width=0.85\textwidth]{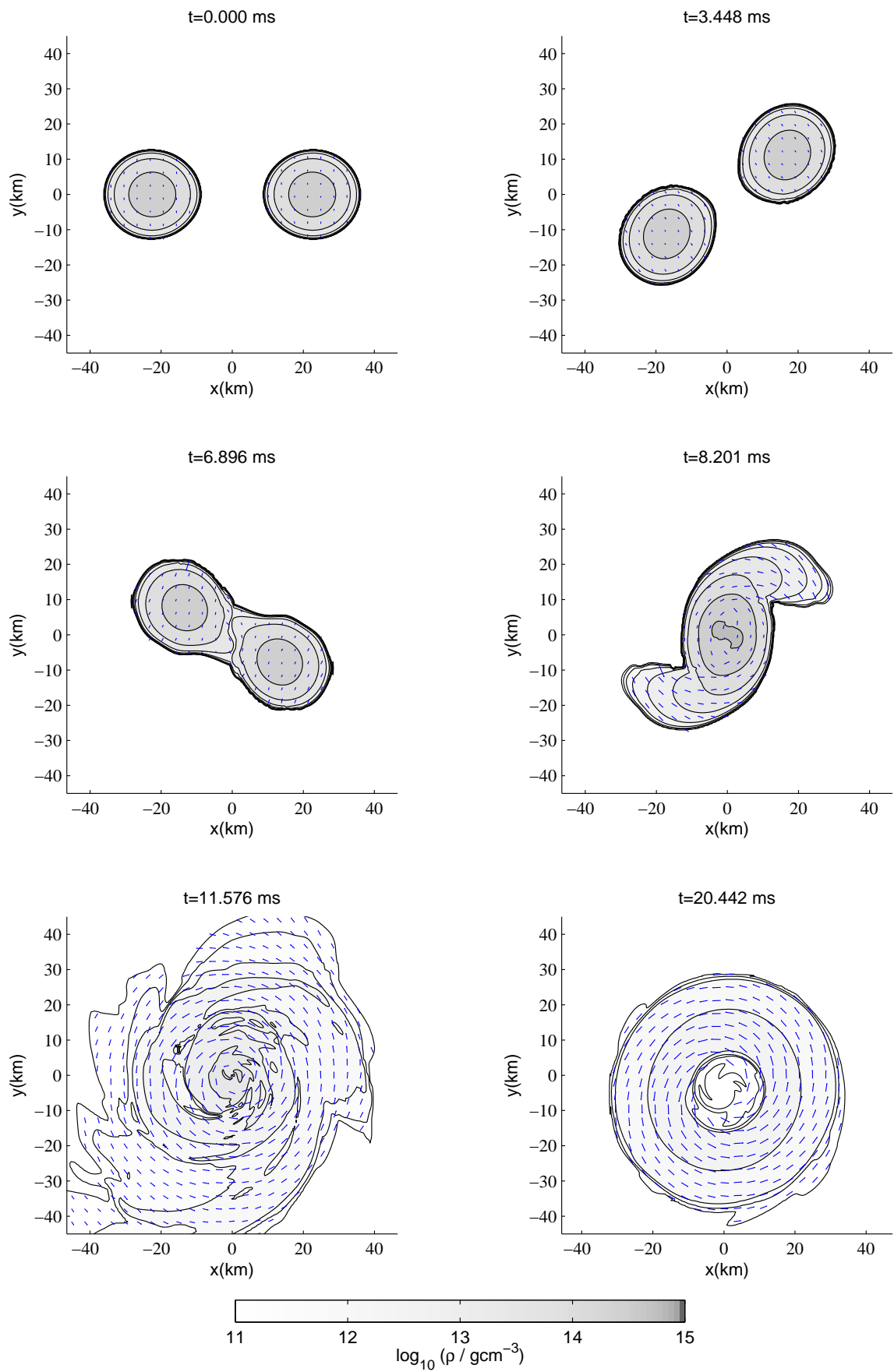}
  \end{center}
  \caption{Dynamics of the G2P14 evolution. The
    picture shows contour plots in the $x-y$ plane of the rest-mass
    density $\rho$ and the velocity $v^i$ at
    different times. 
    Data refer to run H3
    }
     \label{fig:G2P14:ev} 
\end{figure*}

The initial configuration has been evolved using both the cold
polytropic EoS and the ideal gas $\Gamma=2$ EoS.
In the following we will refer to the isentropic evolution with the
name of the initial configuration while we will use the suffix ``hot''
when thermal effects are simulated. Evolutions of this configurations 
were previously performed in~\cite{Baiotti:2008ra,Giacomazzo:2010bx}. 
Overall, we observe a qualitative agreement between our simulations and the
results reported in the literature. Main quantitative differences are
found in the post-merger phase, in particular in the collapse time of the HMNS. 
This is not very surprising since soon after the two stars come in contact the 
convergence of this kind of simulations drops to first order, 
e.g.~\cite{Baiotti:2009gk}. The post-merger phase is thus very dependent on the 
resolution and grid settings employed as well as on the specific HRSC scheme employed. 
In the following we will mostly concentrate our discussion on G2P14 evolutions 
since they were performed at all the grid configurations. The G2P14hot evolutions were
performed up to H1 configuration. Comparison between G2P14hot and
G2P14 are presented for H1. 

Fig.~\ref{fig:G2P14:ev} shows contour plots in the $x-y$ (orbital)
plane of the rest-mass density and the velocity field at different
times of the G2P14 evolution. Data refer to run H3. 
The binary performs about 3 orbits before the merger. We define the
merger time $t_{\rm m}$ as the peak of the (2,2) mode of the GW amplitude, $|h_{22}|$, where
$h\equiv h_+-\i~h_\times$ (see next section). Clearly the two stars 
come in contact before the merger time. 
From the H3 run we have $t_{\rm m}=1765$~($8.69$~ms) while the
contact time is about $1290$~($6.3$~ms). 
After the merger we observe a bar-shaped differentially rotating star
with rest-mass  $\sim2M_{\rm b}$: the HMNS~\footnote{%
  HMNSs are differentially rotating NSs with mass larger than the
  maximum mass of the corresponding (same EoS) stable uniformly
  rotating star. Stable configurations of uniformly rotating stars 
  with mass larger than the maximum allowed for the stability in the
  spherical case do exist, supported by centrifugal forces. They are
  called supra-massive NSs and rotate at the mass-shedding limit. 
}.
Note in Fig.~\ref{fig:G2P14:ev} the initial rotational symmetry of 
the HMNS (obtained without imposing $\pi$-symmetry in the grid) and
how the symmetry is broken during the following dynamics.
The large non-axisymmetric deformation of the HMNS causes a strong GWs 
emission~\cite{Shibata:2000jt,Baiotti:2006wn} which
carries away matter angular momentum.
As a result the HMNS becomes more compact and finally collapses at
about $t_{\rm AH}\sim 2118$~($10.43$~ms) when an apparent horizon is
first formed. A fraction of the matter remains outside the
horizon and forms an accretion disk. Note that in the disk rotational
symmetry is approximately restored. The mass and spin of the BH
estimated from the apparent horizon are rather inaccurate due to a
lack of resolution. They first rapidly grow in time reaching local
maximum, then the BH mass is observed to increase (see
Fig.~\ref{fig:G2P14:BH}) while the angular momentum decreases. We
estimate at the end of the evolution $M_{\rm BH}\sim2.77$ and 
$a_{\rm BH}\equiv J_{\rm BH}/M_{\rm BH}^2\sim0.72$. They have discrepancies
of about 7\% and 25\% with the expected value once the gravitational
radiation emission has been taken into account.  
Note that the collapse simulation discussed in Sec.~\ref{sec:ns:unst},
where the apparent horizon is computed with satisfactory accuracy,
employs in the finest grid level a resolution about ten times higher
than in the H1 runs. Additionally the gauge choice plays an important role in
determining the coordinate size of the BH as discussed below.

The mesh refinement implementation is such that as soon as the
two moving boxes come in contact (which happens before the contact 
of the two stars) a larger box with the same resolution is constructed.
Before the two stars come in touch however it can happen 
that the boxes split back to the initial ones, evolve individually and
merge again. The reason is that the evolution of a very large box is 
computationally not affordable in terms of memory and time.  
This behavior is well tested in BBH
simulations~\cite{Brugmann:2008zz}. It could lead to a lack of
accuracy at the center of the grid but in practice it has a negligible
impact since it happens when the main part of the matter is still
distributed away from the center. 

\begin{figure}[t]
  \begin{center}
    \includegraphics[width=0.48\textwidth]{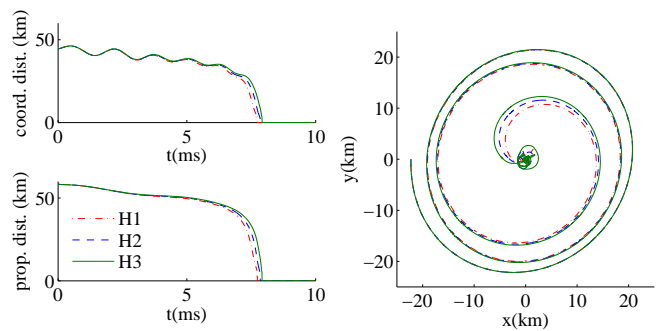}
    \caption{Orbital motion of G2P14.
      The figure shows the evolution of the coordinate (top-left) and
      proper (bottom-left) separation of the binary and the
      star-track (right) of the two stars for different grid settings.}
   \label{fig:G2P14:st} 
  \end{center}
\end{figure}

In the right panel of Fig.~\ref{fig:G2P14:st} we report the star-track
of one of the stars, defined as the minimum of the lapse for
resolutions H1, H2 and H3. The left panel instead shows for the same
resolutions the coordinate (top) and proper (bottom) spatial separation. 
As is evident from the figure, the orbital motion has some eccentricity 
due to the initial data (visible in the proper separation plot) and 
a coordinate eccentricity (visible in the star-tracks and as oscillations 
in the coordinate separation) due to the evolution 
itself. The eccentricity of the initial data is caused by the
conformally-flat approximation; the effect becomes bigger at smaller
separation. In~\cite{Kiuchi:2009jt} the problem is corrected by adding
an ``approaching'' radial velocity based on the post-Newtonian T4 formula. Since
the procedure is constraint violating we prefer not to adopt it.
The contribution to the coordinate eccentricity is mainly
related to the shift condition and discussed below.
The use of a lower resolution results in an earlier
merger. While the truncation errors due to finite resolution are quite
large, the simulations with grid settings H have entered the
``convergent regime'' in the sense that we are able to estimate
convergence. 

\begin{figure}[t]
  \begin{center}
    \includegraphics[width=0.48\textwidth]{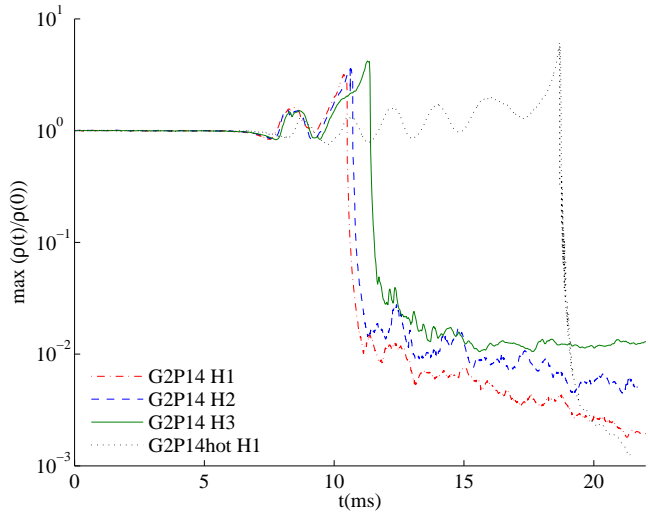}
    \caption{Evolution of the central rest-mass
      density in BNS simulations. Runs G2P14 are reported for
      the three highest resolutions while run G2P14hot for H1.}
    \label{fig:G2P14:rhomax}  
  \end{center}
\end{figure}

The evolution of the maximum rest-mass density is shown in
Fig.~\ref{fig:G2P14:rhomax} for different resolutions. After the
merger about two quasi-radial oscillations of the HMNS are
observed. The maximum density increases by about a factor two
indicating that the HMNS becomes more compact until, finally, it
collapses. The use of more resolution results in a more persistent
HMNS (see below for a comparative discussion with 
G2P14hot). Note the quite large differences between the runs at different resolutions. 
After the collapse the matter of the disk accretes onto the
BH. 

\begin{figure}[t]
  \begin{center}
    \includegraphics[width=0.48\textwidth]{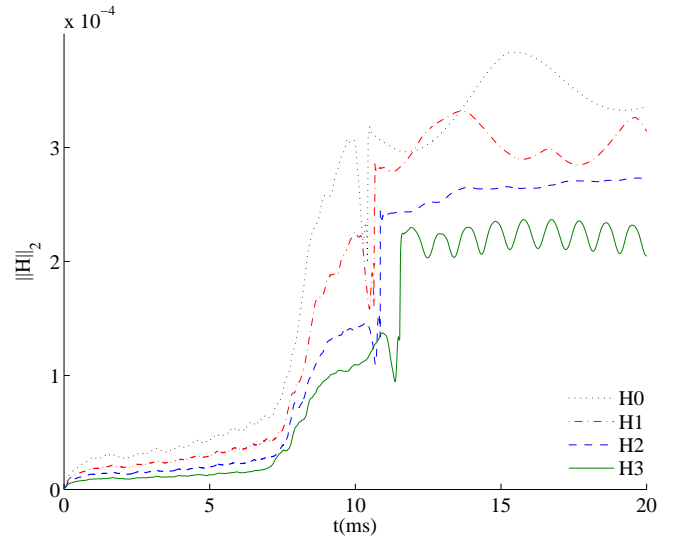}
    \caption{Hamiltonian violation in G2P14
      evolution. The figure shows the evolution of the L2 norm of the
      Hamiltonian constraint for different resolutions.}
    \label{fig:G2P14:Ham} 
  \end{center}
\end{figure}

Concerning the constraints, the biggest violation is found 
in the Hamiltonian. The momentum constraint is generically one order of magnitude smaller 
and becomes comparable to the Hamiltonian constraint only after the formation of the ``puncture''.
The evolution of the L2 norm of the Hamiltonian
constraint is displayed in Fig.~\ref{fig:G2P14:Ham}. The preservation 
of the constraint improves with resolution. We find 2nd
order convergence during the whole inspiral.
Most of the violation is observed in the region covered by the matter,
similarly to what was discussed for the test involving a single star spacetime.
After the two stars come in contact the violation rapidly increases
and the convergence rate drops down to first order. 
At the time when an apparent horizon forms the norms show a large gradient.  

\begin{figure}[t]
  \begin{center}
    \includegraphics[width=0.48\textwidth]{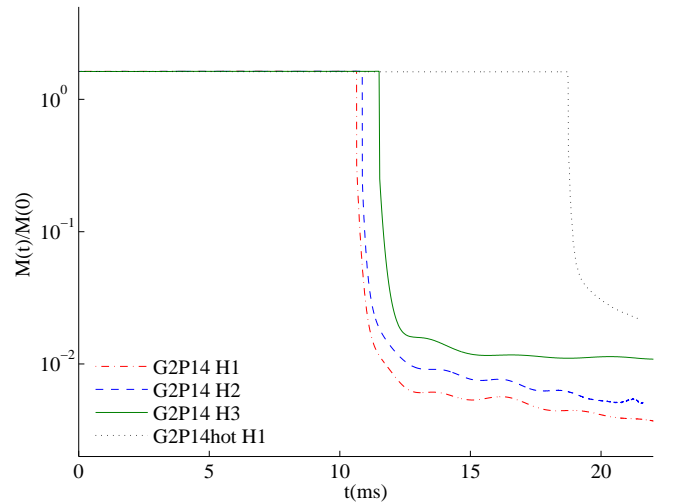}
    \caption{Conservation of rest-mass in BNS simulations.
      The figure shows the evolution of the rest-mass normalized to the initial value for 
      G2P14 and G2P14hot evolutions.
      Runs G2P14 are reported for the three highest resolutions, G2P14hot 
      for resolution H1.
    }
    \label{fig:G2P14:M0}
 \end{center}
\end{figure}

The conservation of the rest-mass is excellent until
the BH forms with a largest deviation of $\Delta
M_0/M_0\sim 1$~\% in the H1 runs and a minimum
of $\Delta M_0/M_0\sim 0.5$~\% in the H3 runs. 
This behavior is illustrated in Fig.~\ref{fig:G2P14:M0} which displays
the evolution of $\Delta M_0/M_0$. The plot shows also matter
accretion after the collapse and indicates an upper limit (from run H3)
for the initial rest-mass of the disk, $M_{\rm d}\lesssim
2\times10^{-2}M_0$, since the integral in Eq.~\eqref{eq:M0} is
performed on the whole grid (including the BH interior).

\begin{figure}[t]
  \begin{center}
    \includegraphics[width=0.48\textwidth]{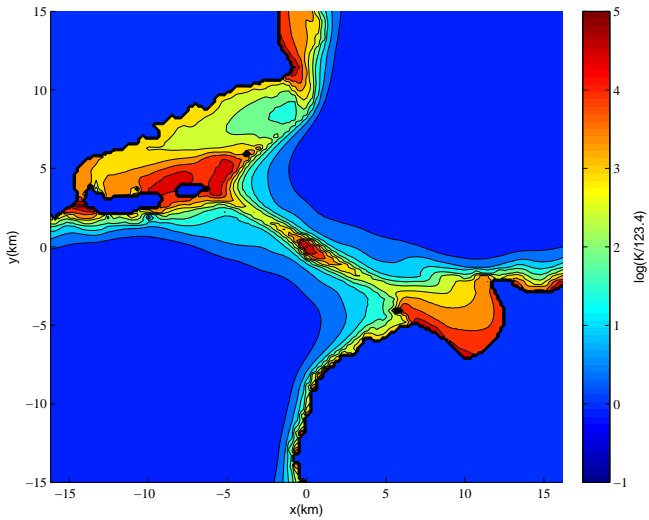}
    \caption{Shock formation in G2P14hot
      evolution. The figure shows the quantity $K=p/\rho^\Gamma$
      normalized by its initial value on the orbital plane at time
      $t=1275$~($6.256$~ms) in a $\log_{10}$ scale.
      Data refer to run H1.}
    \label{fig:G2P14:entropy}  
  \end{center}
\end{figure}

We finally compare the dynamics of G2P14 with G2P14hot. The inspiral
motion of the two binaries present small differences due to spurious
numerical effects. Small errors, triggered by the artificial 
atmosphere treatment, propagate as simulation time advances
and artificially heat the stars especially at their surface. The
effects of these errors on the GWs are quantified in the following
section. When the two stars touch physical effects dominate instead
and the evolutions significantly differ.
Fig.~\ref{fig:G2P14:entropy} shows a contour plot in the
$x-y$ plane of the quantity $K=p/\rho^\Gamma$ (normalized by its
initial value) for the G2P14hot evolution. $K$ gives a simple measure
of the entropy whose variation indicates the presence of a shock 
(see e.g.~\cite{Zanotti:2010xs} for the development of a 
``shock detector''), and in the G2P14 evolution $K$ is constant by construction.
In contrast, during G2P14hot evolution the inclusion of thermal
effects yields shock formation when the arms of the two distorted
stars come in contact. The figure shows exactly this phenomenon 
indicating that the quantity $K$ increases by about 5 orders of
magnitude. The thermal energy of the fluid rapidly increases reaching
peaks of $\epsilon^{\rm hot}\sim 0.025-0.03$ corresponding to
temperatures of $T\sim2\times10^{11}$~K. At the time of the collapse the
average temperature of the fluid is of order $T\sim6\times10^{10}$~K and it
has reached peaks up to $T\sim2\times10^{12}$~K.

Due to the additional pressure support provided by thermal effects (see
Eq.~\eqref{eq:eos_tab1Dhot}) the HMNS in the G2P14hot evolution
collapses later. Fig.~\ref{fig:G2P14:rhomax} indicates that the
collapse takes place after $\sim9$~ms after the HMNS formation. During
this time interval several quasi-radial oscillations at a frequency $\nu_{\rm
  F}\sim473$~Hz and a strong GWs emission are observed (see next
section). 
The central density shows again a drift to larger values
indicating that the star is becoming progressively more compact.
Results similar to G2P14 are found for what concerns rest-mass
conservation as demonstrated in Fig.~\ref{fig:G2P14:M0}.

In~\cite{Baiotti:2008ra} the evolution of G2P14 leads to a prompt
collapse without the formation of a HMNS, while the evolution of
G2P14hot leads to a collapse at about
$t\sim14$~ms (simulations employ the PPM implementation 
described in~\cite{Hawke:2005zw} and the Marquina numerical flux). 
In~\cite{Giacomazzo:2010bx}, where a different grid
setting is employed (but the same code with the HLL numerical
flux), the evolution of G2P14hot leads to a collapse at about
$t\sim16-17$~ms.  
In both works~\cite{Baiotti:2008ra,Giacomazzo:2010bx}, and differently 
from here, the rotational binary symmetry ($\pi$-symmetry) is enforced in the
numerical grid but this procedure is expected at most to lead to a more
persistent HMNS due to non-linear mode couplings (in particular $m=1$
modes)~\cite{Baiotti:2006wn}. It is thus unlikely that the grid
symmetry is the reason for the discrepancy as demonstrated below. 
In the following we show how the HMNS dynamics is very sensitive to the 
numerical method employed presenting results from simulations that employ 
different reconstruction methods but are otherwise equivalent.
Together with the low convergence of the HRSC methods in presence of
turbulence~\cite{Baiotti:2008ra} or shocks and given a dependence on
the grid settings~\cite{Baiotti:2009gk} the differences observed are
not surprising.

\paragraph*{Effect of reconstruction methods.} 
As pointed out in Sec.~\ref{sec:ns} the different
reconstruction methods have different truncation error magnitudes
and once inserted in the algorithm produce quantitatively
different results. Ref.~\cite{Giacomazzo:2009mp} already pointed out 
that the use of very dissipative limiters such as MM2 can lead to very 
different waveforms with respect to (formally) more accurate methods 
while maintaining the same nominal order of convergence. 
Our simulations confirm this conclusion and here
we show how dramatic the influence of a dissipative scheme on
the dynamics can be.

\begin{figure}[t]
  \begin{center}
    \includegraphics[width=0.48\textwidth]{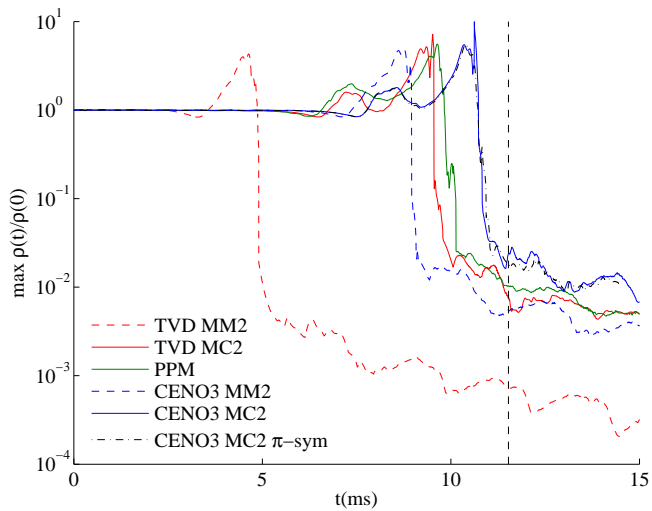}
    \caption{Dependence of the dynamics
      on the use of different reconstruction schemes. The
      figure shows the maximum of the rest-mass density obtained from
      G2P14 evolutions with different
      reconstruction methods.
      Reference vertical line corresponding to $\max\rho$ of run H3.
    }
    \label{fig:G2P14:rhomax_rec} 
  \end{center}
\end{figure}

We consider G2P14 evolutions at resolution H1 with the following
reconstructions: MM2, MC2, PPM, CENO and CENO based on the MM2
sub-stencil (instead of the standard MC2~\cite{Zanna:2002qr}). Since
the resolution is not optimal we expect the effects due to different
truncation errors to be strongly
magnified. Fig.~\ref{fig:G2P14:rhomax_rec} shows the evolution of
the maximum rest-mass density for the different methods. The vertical
line marks the maximum of $\rho$ computed from the H3 run. 
A visual inspection of the data is already conclusive in this case. As  
partially expected, the most diffusive schemes lead to an earlier
merger and collapse while formally high-order reconstructions are
closer to our best estimate. In the simulation with MM2 
reconstruction the inspiral is very short, the HMNS is not formed and
the binary evolution results in an early prompt collapse. 
The simulations with PPM and MC2 give similar results slightly
speeding up the collapse with respect CENO.
When we introduce a more dissipative component in the CENO
limiter, i.e.\ the MM2 linear sub-stencil, the global results are 
significantly affected: an earlier prompt collapse is observed.
The specific reason is obviously the fact that the linear sub-stencil
MM2 is often selected by the limiter. We mention here that, according
to our experience, MM2 reconstruction does not 
guarantee the long-term-stability of an equilibrium spherical star even
in 1D~\cite{Bernuzzi:2009ex}. 

Clearly the big differences shown here become progressively smaller
when higher resolutions are considered. However, due to the slow (2nd)
order convergence of HRSC, we expect they play a major role also at the
higher resolutions employed for standard production runs.  
We point out that in comparing some runs with grid configurations L, M
and H0 we observed a non-monotonic behavior of the results for  
increasing resolution. This effect seems to be related to the poor
resolution but we can not exclude that the phenomenon will also happen
at higher resolutions. Finally, the figure demonstrates that the use of
$\pi$-symmetry does not have a relevant role (at least at this
resolution) in the persistence of the HMNS as mentioned above.

\paragraph*{Gauge study: the $\eta$ parameter.} 
Among the parameters entering the gauge conditions in 
Eq.~\eqref{eq:lapse_1log} and Eq.~\eqref{eq:beta_Gdriv}
the damping parameter $\eta$ in the Gamma-driver shift equation 
has been recently the subject of investigation for BBH
simulations with unequal masses~\cite{Muller:2009jx,Lousto:2010tb,Lousto:2010qx,Lousto:2010ut,Schnetter:2010cz,Muller:2010zze,Alic:2010wu}.
In case of equal mass BBHs it is typically set to $\eta=1/M$ or $\eta=2/M$, where
M is the sum of the ADM masses of the two holes ($M=1$), and it has been
proved to be important to properly resolve the holes on the coordinate 
grid~\cite{Brugmann:2008zz}. 
In the case of BNSs it has been suggested~\cite{Baiotti:2010ka} that 
results are not significantly affected by this choice. Different values 
are adopted in the literature,
e.g.~$\eta=3/M_{\rm b}$~\cite{Yamamoto:2008js,Baiotti:2010ka},
$\eta=1$~\cite{Baiotti:2008ra}, without a detailed analysis. 
We demonstrate in the following that in our set up different choices
of $\eta$ do have an influence on the dynamics of the inspiral. In
particular smaller values of $\eta$ lead to a less eccentric coordinate 
orbital motion and to a smaller coordinate size of the final BH. 

\begin{figure}[t]
  \begin{center}
    \includegraphics[width=0.48\textwidth]{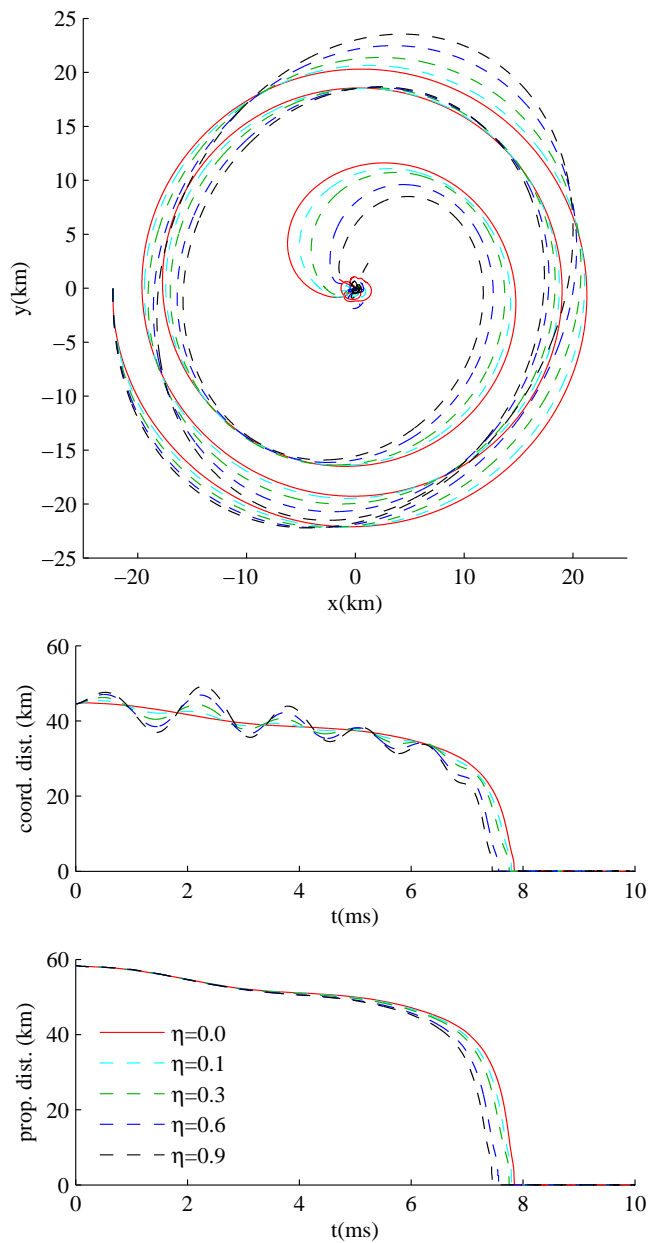}
    \caption{ Effect on orbital dynamics of
      different choices of the parameter $\eta$ in the Gamma-driver
      shift condition. The top panel shows the star-track of one
      star. The central panel shows the evolution of the
      coordinate separation while the bottom panel shows the evolution
      of the proper separation of the binary.  
      Each line refers to a different G2P140 evolution with a
      different $\eta$. Data refer to H1 runs.}
    \label{fig:G2P14:eta} 
  \end{center}
\end{figure}

We consider G2P14 simulations with grid configuration H1 and values 
$\eta=0,\,0.1,\,0.3,\,0.6,\,0.9,\,1.8$,
i.e.~$\eta\simeq0/M,\,0.3/M,\,0.9/M,\,1.8/M\,2.7/M,\,5.4/M$.
Fig.~\ref{fig:G2P14:eta} shows the star-track (top panel), the 
evolution of coordinate separation (central panel), and of the proper
separation (bottom panel) from simulations with different values of
$\eta$. The coordinate separation shows large oscillations and a systematic
shift of the merger to earlier times for higher values of $\eta$.  
The latter is also visible in the evolution of the proper separation, which
instead highlights non-circular effects due to the conformally flat initial data.
The star-track shows how the coordinate eccentricity progressively
reduces when smaller values of $\eta$ are employed. 
When considering the gravitational wave emission the coordinate eccentricity 
does influence the extracted waves and it manifests itself in a phase difference
accumulating during the inspiral. 
Taking as a reference the $\eta=0$ case, the difference in
the phase $\phi$ of $r\,\psi^4_{22}$ (see Sec.~\ref{sbsec:bns:waves}) 
computed with $\eta=0$ is of
$\Delta\phi=0.43,\,0.90,\,2.25,\,3.20$~rad respectively for
$\eta=0.1,\,0.3,\,0.6,\,0.9$ at the merger time. 

\begin{figure}[t]
  \begin{center}
    \includegraphics[width=0.48\textwidth]{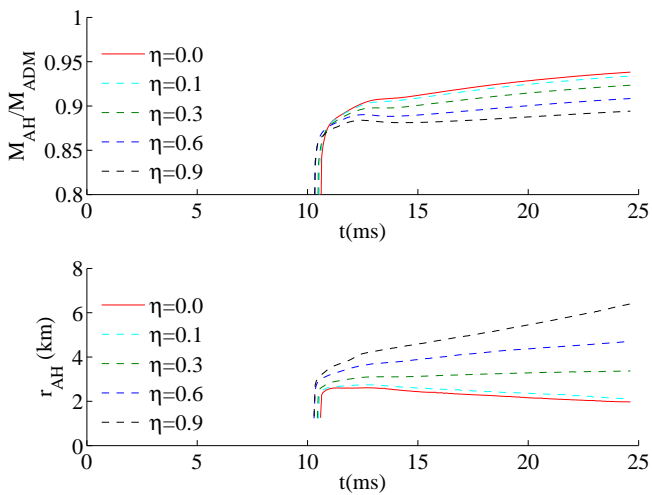}
    \caption{Effect on the final BH mass
      and coordinate radius of the parameter $\eta$ in the Gamma-driver
      shift condition. The top panel shows the irreducible mass
      of the final BH normalized by the ADM mass of the system as
      computed from the apparent horizon finder. The bottom panel
      shows the coordinate radius of the apparent horizon.
      Each line refers to a different G2P140 evolution with a
      different $\eta$. Data refer to H1 runs.}
    \label{fig:G2P14:BH} 
  \end{center}
\end{figure}

Fig.~\ref{fig:G2P14:BH} shows the irreducible mass, $M_{\rm AH}$
(normalized to the ADM mass of the system), and the coordinate radius,
$r_{\rm AH}$, of the apparent horizon. As expected the coordinate
size of the final BH is larger for higher values of $\eta$. 
The choice of $\eta$ must be guided by the requirement of minimizing
the coordinate eccentricity of the orbital motion while keeping the ability of 
correctly resolving the final BH on the finest grid level. 
The use of $\eta=1.8$ for example is not shown in the
figure because it produces an unacceptable eccentricity in the dynamics (see
Fig.~\ref{fig:G2P14:eta}) and the final BH has a size not
completely contained in the refinement level 5, resulting in very
inaccurate results. 
In a similar way the use of large $\eta$ also affects the coordinate radius of the
GW extraction spheres; smaller proper radii corresponds to higher $\eta$.

\subsection{Gravitational waves}
\label{sbsec:bns:waves}

Gravitational radiation plays the fundamental role driving the 
dynamics of the binary system. GWs encode information from each 
phase of the evolution from the inspiral to the collapse and the 
BH-disk formation. We analyze in the following the 
GWs computed from our simulations. Again we focus the discussion on  
G2P14 evolution.  

\begin{figure}[t]
  \begin{center}
    \includegraphics[width=0.48\textwidth]{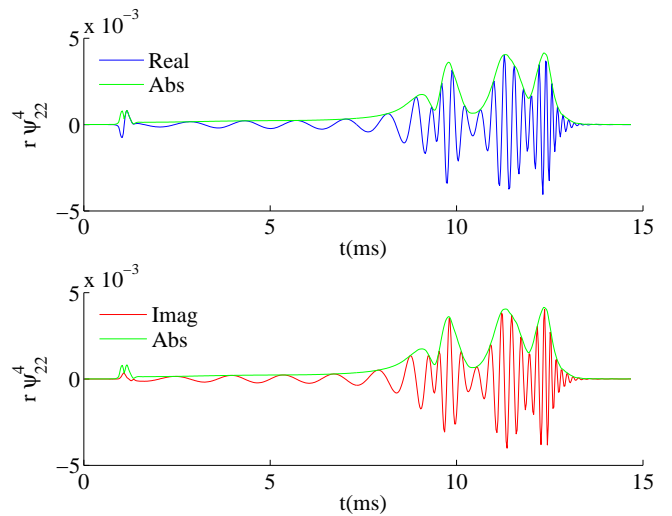}
    \caption{$\psi_4$ waveforms from
      G2P14 evolution. The figure shows the real part (top panel) and
      the imaginary part (bottom panel) of the $r\,\psi^4_{22}$ waveform
      extracted at $r=200$ from the H3 run. The amplitude is
      also shown in both panels.}
    \label{fig:G2P14:rpsi4} 
  \end{center}
\end{figure}

Our simulations show that about $1\,\%$ of the initial ADM energy and 
$13\,\%$ of the initial angular momentum are emitted in GWs during the 
simulation. 
The main emission channel is the $(\ell,m)=(2,2)$ mode of the
multipolar ($s=-2$ spin-weighted spherical harmonics) decomposition of
GWs. The $(2,2)$ mode includes about $97$~\% of the entire radiated
energy, thus we focus on the analysis of this
mode. Fig.~\ref{fig:G2P14:rpsi4} shows the complex $r\,\psi^4_{22}$
waveform computed from the Newman-Penrose scalar $\psi_4$ as described
in~\cite{Brugmann:2008zz} from run H3. GWs are extracted on coordinate
spheres on grid level 1.
The complex waveform is usually decomposed in amplitude and phase, 
$r\,\psi^4_{22} = A \exp(-\i \phi)$. Fig.~\ref{fig:G2P14:rpsi4} shows 
the real (blue solid line) and the imaginary part (red dashed line) as
well as the amplitude (green solid line). The extraction sphere is
placed at coordinate radius $r=200$~($295.3$~km). We checked that
waves extracted at different radii  $r\geq100$ show the proper 
fall-off behavior.
The waveform is plotted against the simulated time instead of the more
appropriate retarded time, see below. 
We present metric waveforms only with respect to the retarded time.
From the plot one clearly identifies the inspiral phase followed by
the emission related to the HMNS oscillations and then the collapse. 
At early times the well known initial ``junk'' radiation can be seen. 

\begin{figure}[t]
  \begin{center}
    \includegraphics[width=0.48\textwidth]{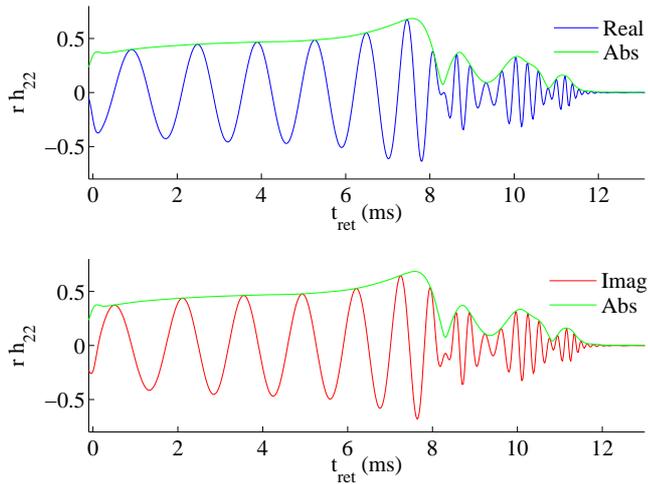}
    \caption{Metric waveforms from 
      G2P14 evolution. The figure shows the real part (top panel) and
      the imaginary part (bottom panel) of the $r\,h_{22}$ waveform
      extracted at $r=200$ from H3 run. The amplitude is
      also shown in both panels. The waveforms are plotted
      versus the retarded time $t_{\rm ret}\equiv t -r_*$ where $r_*$
      is the tortoise radius corresponding to $r$. The metric waveform
      is computed with the FFI and cutoff frequency $\nu_0=0.002$.}
    \label{fig:G2P14:h}
  \end{center}
\end{figure}

\paragraph*{Metric waveforms.} The $\psi_4$ waveform is the second
derivative of the metric waveform $h\equiv h_+-\i~h_\times$, which
represents the actual GW degrees of freedom. The integration of the
relation
\begin{equation}
\label{eq:psi2h}
\ddot{h}_{\ell\,m} = \psi^4_{\ell\,m} \  
\end{equation}
to obtain $h$ multipoles from those of $\psi_4$, requires
some attention. 
In~\cite{Pollney:2007ss,Berti:2007fi,Baker:2008mj,Damour:2008te,Baiotti:2008nf}
$h$ is computed via a direct (time domain) integration on the
simulated time domain. The result is affected by a polynomial drift 
that must be corrected by fitting. We refer to this procedure as the
corrected time domain integration (CTI).  
The drift observed in the raw integration is not only the linear
contribution expected from the integration of Eq.~\eqref{eq:psi2h} but a
generic polynomial. It originates from the integration of
high-frequency noise in the data and has a stochastic
nature~\cite{Reisswig:2010di}.  
Recently~\cite{Baiotti:2011am} an improved CTI procedure, which
amounts to subtracting a post-Newtonian behavior before fitting the
polynomial correction, which we did not consider here, has been developed. 
In~\cite{Reisswig:2010di} the fixed-frequency integration (FFI)
method has been proposed. It is based on a spectral integration
in the Fourier basis and employs a simple high-pass frequency filter 
against spectral leakage.

We computed $h_{22}$ with both the CTI and the FFI procedure finding
comparable results. Differences are discussed below. We focus for the
moment on $h_{22}$ computed with FFI: the integration method does not
influence the following statements. Fig.~\ref{fig:G2P14:h} shows the
metric waveform $r\,h_{22}$ computed from $r\,\psi^4_{22}$ as a
function of the retarded time $t_{\rm ret}\equiv t -r_*$ where $r_*\simeq222$
is the tortoise radius~\footnote{ %
  The tortoise radius is computed as $r_*=r_{\rm Schw} + 2M
  \log(r_{\rm Schw}/(2M)-1)$, where $r_{\rm Schw}$ is the
  Schwarzschild radius corresponding to the isotropic radius $r$.   
  The retarded time based on this coordinate, $t_{\rm ret}=t-r_*$, is 
  a useful but approximate quantity which becomes rigorous only at
  large radii when the spacetime becomes Schwarzschild.
  The correspondence dynamics-waveform in the simulation is thus
  biased by this approximation.
}
corresponding to $r=200$. Real part, imaginary part
and amplitude are shown for run H3. The peak of the amplitude of
$h_{22}$ formally defines the merger time, $t_{\rm m}$,
e.g.~\cite{Bernuzzi:2010xj,Baiotti:2011am}. 
Considering different resolutions we found  
$t_{\rm m}=1670,\, 1710,\, 1740$ and $1765$~($8.23$, $8.42$, $8.57$ and
$8.69$~ms), respectively, for runs H0-3. The corresponding retarded
times are $t_{\rm m,
  ret}=1446,\,1486\,1516,\,1541$~($7.12,\,7.32,\,7.47,\,7.59$~ms). The
metric waveform is composed at early times of six GW cycles emitted 
during the three orbit inspiral. After the merger the emission is
dominated by the bar-deformed HMNS, and the signal has a 
typical frequency around $3$~kHz which increases as the HMNS becomes
more compact~\cite{Bernuzzi:2008fu,Manca:2007ca}. Finally, after
$t_{\rm ret}>2132$~($10.5$~ms) the GW signal is composed of the
quasi-normal-mode ringing of the BH. 
We observe from the waves the fundamental
frequency $\nu_{\rm QNM} \sim 6.5 $~kHz, compatible with the estimate
of the BH mass and spin from the apparent
horizon~\cite{Echeverria:1989hg,Baiotti:2008ra}, i.e.\ $\nu_{\rm
  QNM}\sim3.23(10/M_{\rm BH})[1-0.63(1-a_{\rm BH})^{0.3}]$. 
The value $\nu_{\rm QNM}$ can be estimated by fitting the 
plateau of the GW frequency when it reaches its absolute maximum  (see 
Fig.~\ref{fig:G2P14:gwf}). A better estimate is provided by the
frequency of the $\psi_4$ waveforms because the signal is less noisy
and not contaminated by the integration procedure. 

\begin{figure}[t]
  \begin{center}
    \includegraphics[width=0.48\textwidth]{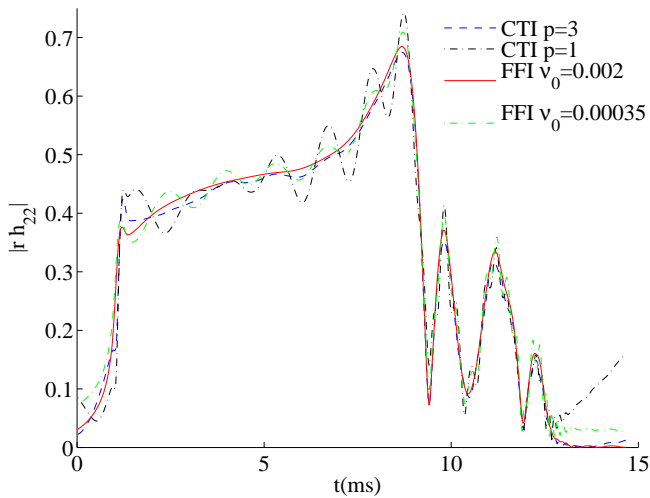}
    \caption{Gravitational wave amplitude
      $|r\,h_{22}|$ from G2P14 evolution. The figure shows 
      the amplitude computed with the FFI $\nu_0=0.002$ (red solid
      line) and $\nu_0=0.00035$ (green dashed-dotted line) and with
      the CTI where a cubic (blue dashed line) and
      linear polynomial (black dotted line) correction is used.}
      \label{fig:G2P14:ah} 
  \end{center}
\end{figure}

We comment now on the two integration methods employed in the
post-processing of $\psi_4$ to obtain $h$. In the FFI we use as
cut off frequency the value $\nu_0=0.002$~($406$~Hz), slightly below the
GW frequency of the initial data. 
The polynomial correction employed
in the CTI is a 3rd order polynomial. This choice is
preferred against a linear or quadratic correction since 
it minimizes experimentally the drift in the raw integrated waveforms 
and the oscillations in the modulus (see below). 
The phase difference between the two metric waveforms amounts to 
$\Delta\phi\lesssim0.06$ from early times until the collapse. 
This value is very small and can be considered insignificant (see 
error estimates below and Tab.~\ref{tab:gws:errors}). On the other
hand the amplitudes do differ in a relevant way. 
The relative difference is less than $\Delta A/A<5$\% during the
inspiral and grows to about 15\% before the collapse. While these
numbers do not seem dramatic another important effect is observed.
Fig.~\ref{fig:G2P14:ah} shows the amplitude of $r\,h_{22}$
computed with the two methods: solid red line for FFI and dashed blue line
for CTI. The amplitude computed with the CTI shows oscillations during
the inspiral. These oscillations converge away considering higher
resolutions but we were not able to remove them completely
(the oscillation amplitude is larger in waveforms from lower
resolutions runs). Waves extracted at larger extraction radii show smaller oscillations. 
The use of a linear polynomial correction results
in very large oscillations (black dotted line) and a prominent drift
in the waveforms. By contrast the
amplitude obtained by the FFI waveform integrated with $\nu_0=0.002$
is free from these oscillations for all the resolutions
considered. When the FFI with a much lower cutoff frequency,
$\nu_0=0.00035$~($71$~Hz), is employed the oscillations appear again
(green dashed-dotted line). Note that a frequency of $71$~Hz roughly 
corresponds to the finite length of the signal.
As observed in~\cite{Damour:2008te,Reisswig:2010di} the
oscillations are an unphysical effect not related to eccentricity.
We mention however that we find here a correlation between the oscillations in the 
amplitude of $r\,h_{22}$ and those seen in the coordinate separation (Fig.\ref{fig:G2P14:eta})  
in the runs with different values $\eta$.
A gauge effect on the extraction spheres amplified in the calculation of $h$ is thus a 
possible explanation.
A proper choice of $\nu_0$ in the FFI can mostly eliminate them, 
while polynomial fitting in CTI is not as robust and not performing as well for our data.

\begin{figure}[t]
  \begin{center}
    \includegraphics[width=0.48\textwidth]{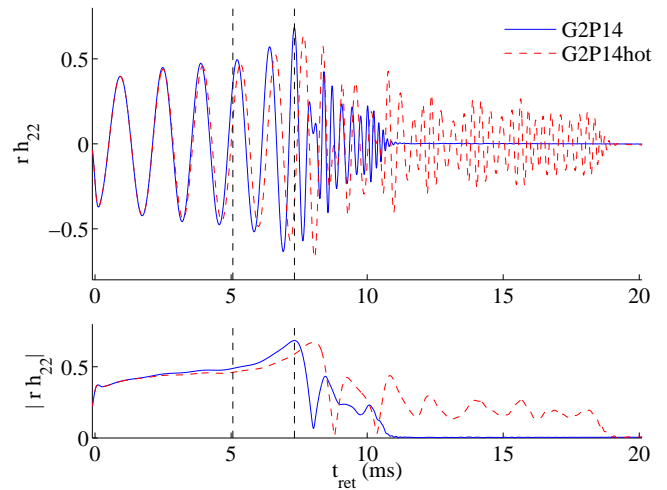}
    \caption{ Comparison between
      waveforms from G2P14 and G2P14hot evolutions.
      The top panel shows the real part of $h_{22}$ extracted
      at $r=200$ for the two evolutions. The bottom panel shows the
      amplitudes. Vertical lines mark the time of the first contact
      between the stars and the merger time for G2P14 evolution. The
      waveforms are plotted against the retarded time $t_{\rm
        ret}\equiv t -r_*$. Data refer to run H1.}
    \label{fig:G2P14:rpsi4_therm}
  \end{center}
\end{figure}

\begin{figure}[t]
  \begin{center}
    \includegraphics[width=0.48\textwidth]{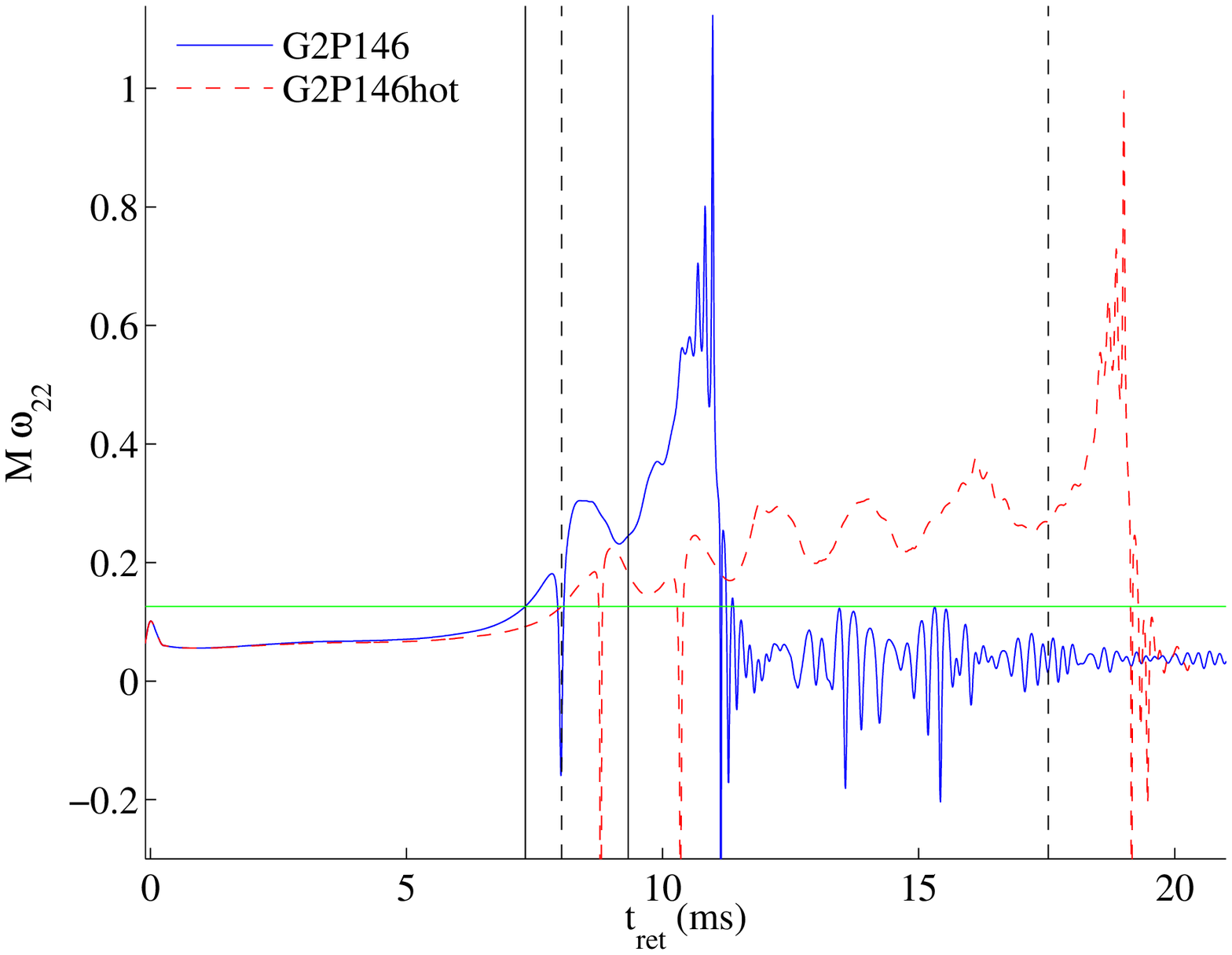}\\
    \includegraphics[width=0.48\textwidth]{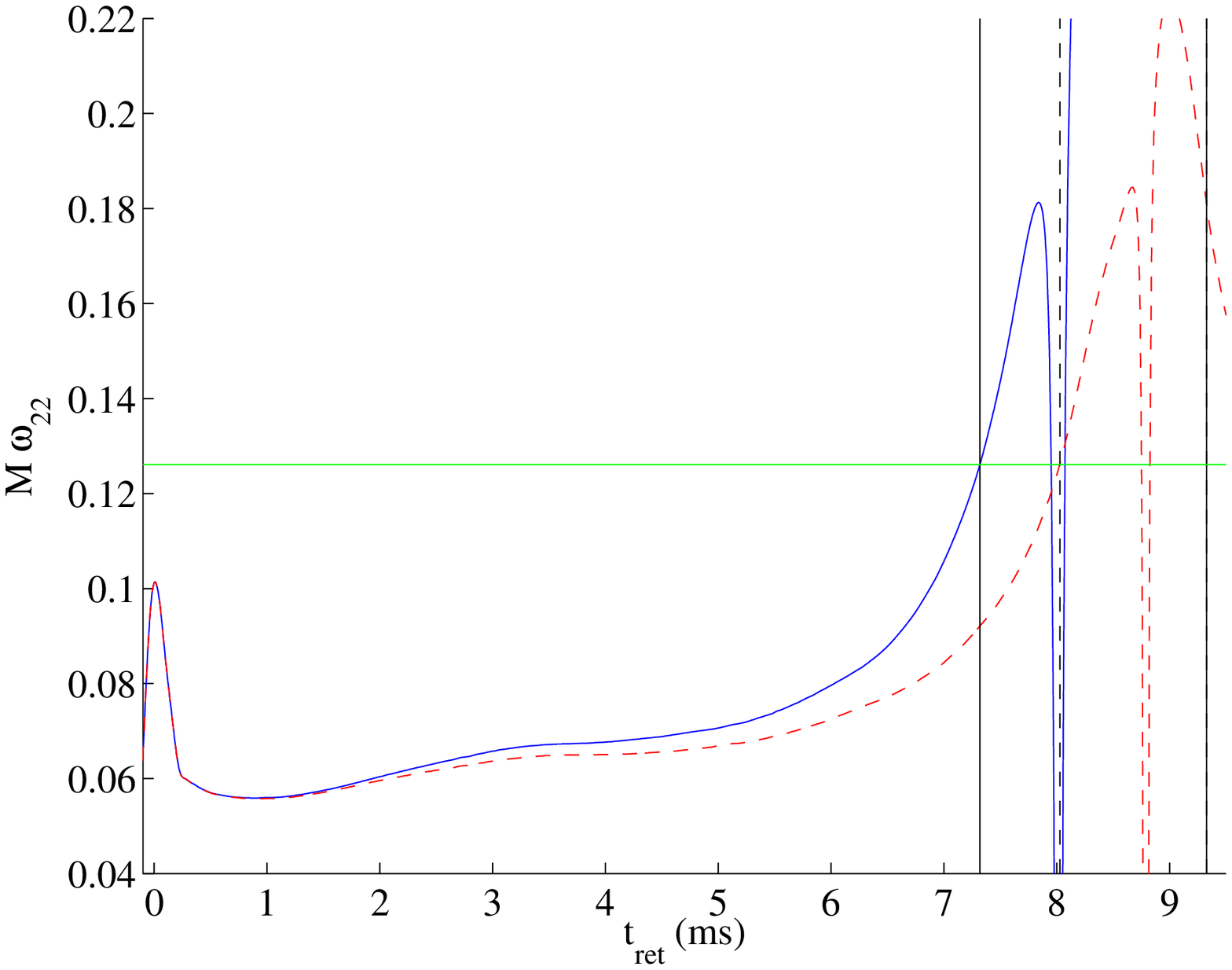}
    \caption{ Gravitational wave frequency from
      G2P14 and G2P14hot evolutions.  
      The figure shows $M\,\omega_{22}$ computed from $h_{22}$
      waveforms. In the top panel the blue solid line refers to G2P14
      and the red dashed line to G2P14hot. The vertical black solid
      lines mark respectively the merger time and the apparent horizon
      formation for G2P14. The vertical black dashed
      lines mark respectively the merger time and the apparent horizon
      formation for G2P14hot. 
      The bottom panel gives a detail of the top panel showing the inspiral part.
      The horizontal green solid line marks the value of
      $M\,\omega_{22}$ at the merger for G2P14. 
      Data refer to run H1.}
    \label{fig:G2P14:gwf} 
  \end{center}
\end{figure}

\paragraph*{Thermal effects.} We discuss now differences between the
waveforms produced in G2P14 and G2P14hot evolution. The results
refer to H1 runs and they are discussed referring to 
Fig.~\ref{fig:G2P14:rpsi4_therm}. As discussed in
Sec.~\ref{sbsec:bns:dyn} the simulations show differences of numerical 
origin already in the inspiral. The phase difference between the
$h_{22}$ waveforms increases monotonically during the evolution until
$t_{\rm ret}\sim12$~ms when the emission of G2P14 is practically zero. At the
time the two stars experience the first contact the G2P14hot $h_{22}$ waveform
has accumulated a dephasing of $\Delta\phi=+0.67$~rad and the amplitude is
about $6$~\% smaller. Since these differences are compatible with the 
truncation errors for this resolution and they are not expected in the
continuum limit (the inspiral is an isentropic process from the fluid
point of view) they likely have a numerical origin.  
After the contact, thermal effects drive a very different dynamics compared 
to the isentropic case as discussed in Sec.~\ref{sbsec:bns:dyn}
(see Fig.~\ref{fig:G2P14:entropy}).  
The dephasing reaches $\Delta\phi=+2.11$~rad at the merger time and
the amplitude is about $14$~\% smaller.
Similarly for the $r\,\psi^4_{22}$ waveform we found a dephasing of
$\Delta\phi=+0.67$~rad and a factor $-15$~\% in amplitude at the
contact time and a dephasing of $\Delta\phi=+2.43$~rad and a factor
$-50$~\% in amplitude at the merger time.
The post-merger emission in the G2P14hot evolution is dominated by
the emission of the bar-deformed HMNS. A lower frequency modulation of
the signal is visible in the amplitude of the waveform and corresponds
to the nonlinear quasi-radial pulsations shown in
Fig.~\ref{fig:G2P14:rhomax} and previously discussed. The
high-frequency part of the GW signal is instead dominated by $m=2$
non-axisymmetric nonlinear modes and it is composed of few frequencies
around $2.7$~kHz. A Fourier analysis shows results qualitatively compatible with~\cite{Stergioulas:2011gd}.

A relevant quantity directly connected to the dynamics of the system
is the GW frequency defined as
$\omega_{22}\equiv-\Im(\dot{h_{22}}/h_{22})$, i.e.\ the derivative of
the GW phase. Fig.~\ref{fig:G2P14:gwf} displays 
the dimensionless $M\,\omega_{22}$ as computed from G2P14 and G2P14hot
evolutions. During the inspiral the GW frequency increases
monotonically from $M\,\omega_{22}\sim0.056$ to
$M\,\omega_{22}\sim0.126$. This value (green horizontal solid line,
see the bottom panel) is common to both evolutions but occurs at different
times due to the accumulated phase. After the merger, a first local
maximum is observable with again very close values for G2P14
($M\,\omega_{22}\sim0.181$) and G2P14hot ($M\,\omega_{22}\sim0.184$)
and different times ($t_{\rm ret}\sim7.84$~ms for G2P14hot and
$t_{\rm ret}\sim8.68$~ms for G2P14hot). The GW frequencies present the same 
behavior until this point, thermal effects generate only a time shift (retardation).
After the merger, the GW frequency of G2P14
presents one oscillation and increases from $M\,\omega_{22}\sim0.2$ to
$M\,\omega_{22}\sim0.245$ when the apparent horizon forms. With a
steep gradient it further increases to $M\,\omega_{22}\sim0.57$ which
correspond to the fundamental QNM of the Kerr BH ($\nu_{\rm QNM} \sim 6.5 $~kHz). 
In the case of G2P14hot the GW frequency reflects the dynamics of the
HMNS: it increases almost linearly with large oscillations
corresponding to the HMNS quasi-radial oscillations. After the local
minimum of the last oscillation the collapse happens at a frequency
$M\,\omega_{22}\sim0.27$.
The QNM frequency on the G2P14hot evolution is slightly below the
corresponding isentropic evolution, $\nu_{\rm QNM} \sim 6.45$~kHz. 
Interestingly the GW frequency becomes negative after the first local
maximum after the merger in both evolutions and one oscillation later
in G2P14hot (after the formation of the apparent horizon in the
G2P14 evolution). It is difficult to say if these features have a
physical or a numerical origin. However, the GW frequency zeros  
and negative spikes, roughly correspond to times at which the 
distribution of the rest-mass density in the equatorial plane is
almost-spherical. 
Note in addition that the amplitude of the GWs drops 
to zero (bottom panel of Fig.~\ref{fig:G2P14:rpsi4_therm}).

\begin{figure}[t]
  \begin{center}
    \includegraphics[width=0.48\textwidth]{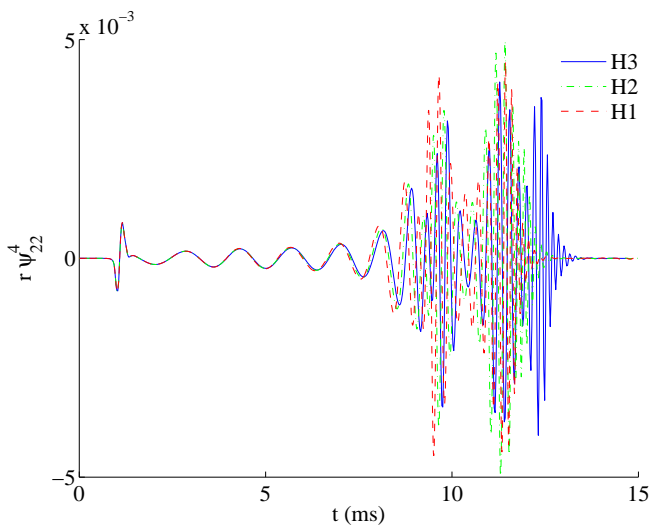}
    \caption{$\psi_4$ waveforms from
      G2P14 evolution at different resolutions. 
      The figure shows the real part of the $r\,\psi^4_{22}$ waveform
      extracted at $r=200$ from runs H1-3.}
      \label{fig:G2P14:rpsi4_diffres}
  \end{center}
\end{figure}

\paragraph*{Convergence.} We discuss now convergence of the waveforms  
produced in G2P14 evolutions and runs H1, H2 and H3. The real part of
$r\,\psi^4_{22}$ is reported in Fig.~\ref{fig:G2P14:rpsi4_diffres}
for the different runs. The convergence study is focused on the
inspiral part of the wave. Similar results are found for the series
H0, H1 and H2, with larger absolute errors.   

\begin{figure}[t]
  \begin{center}
    \includegraphics[width=0.48\textwidth]{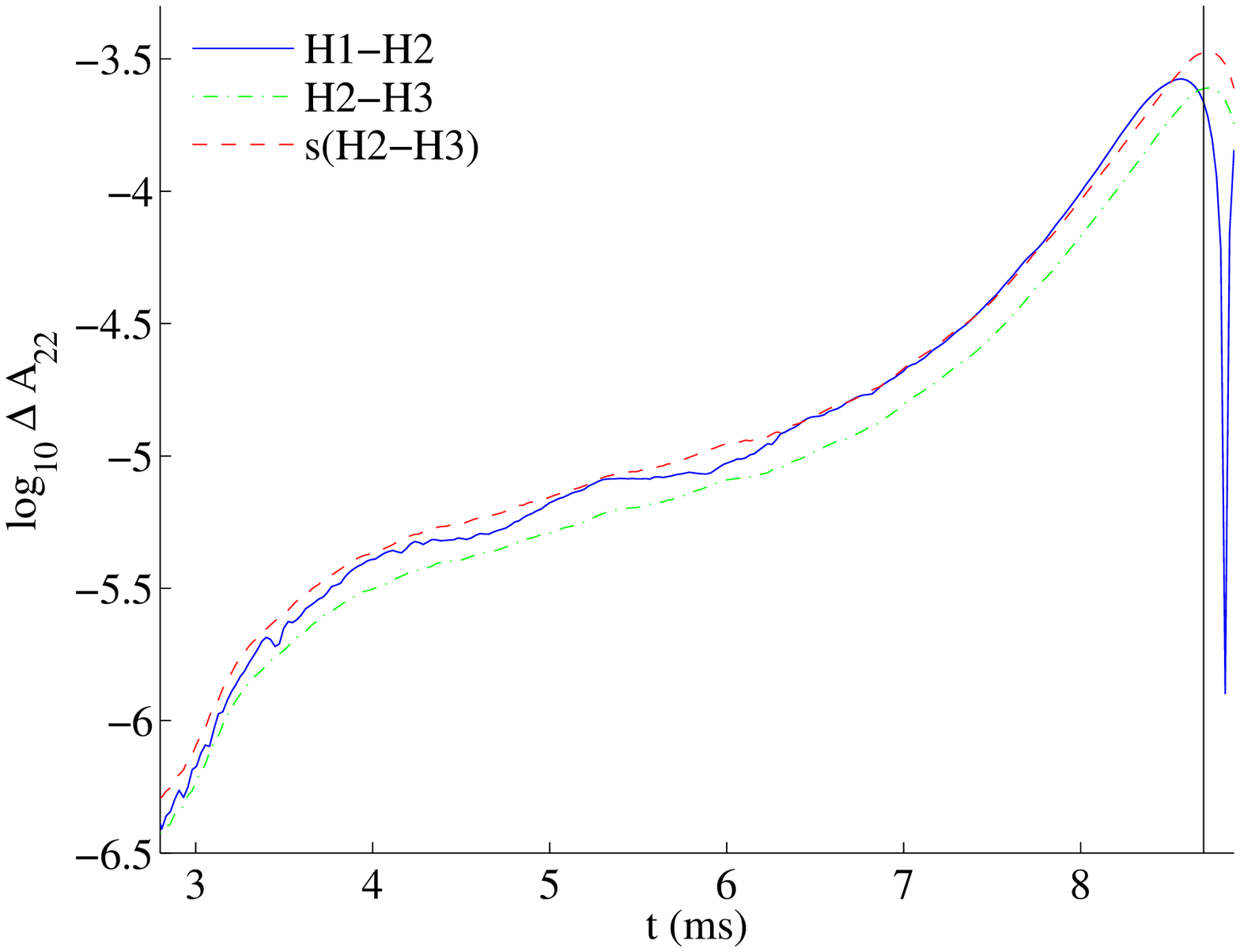}
    \includegraphics[width=0.48\textwidth]{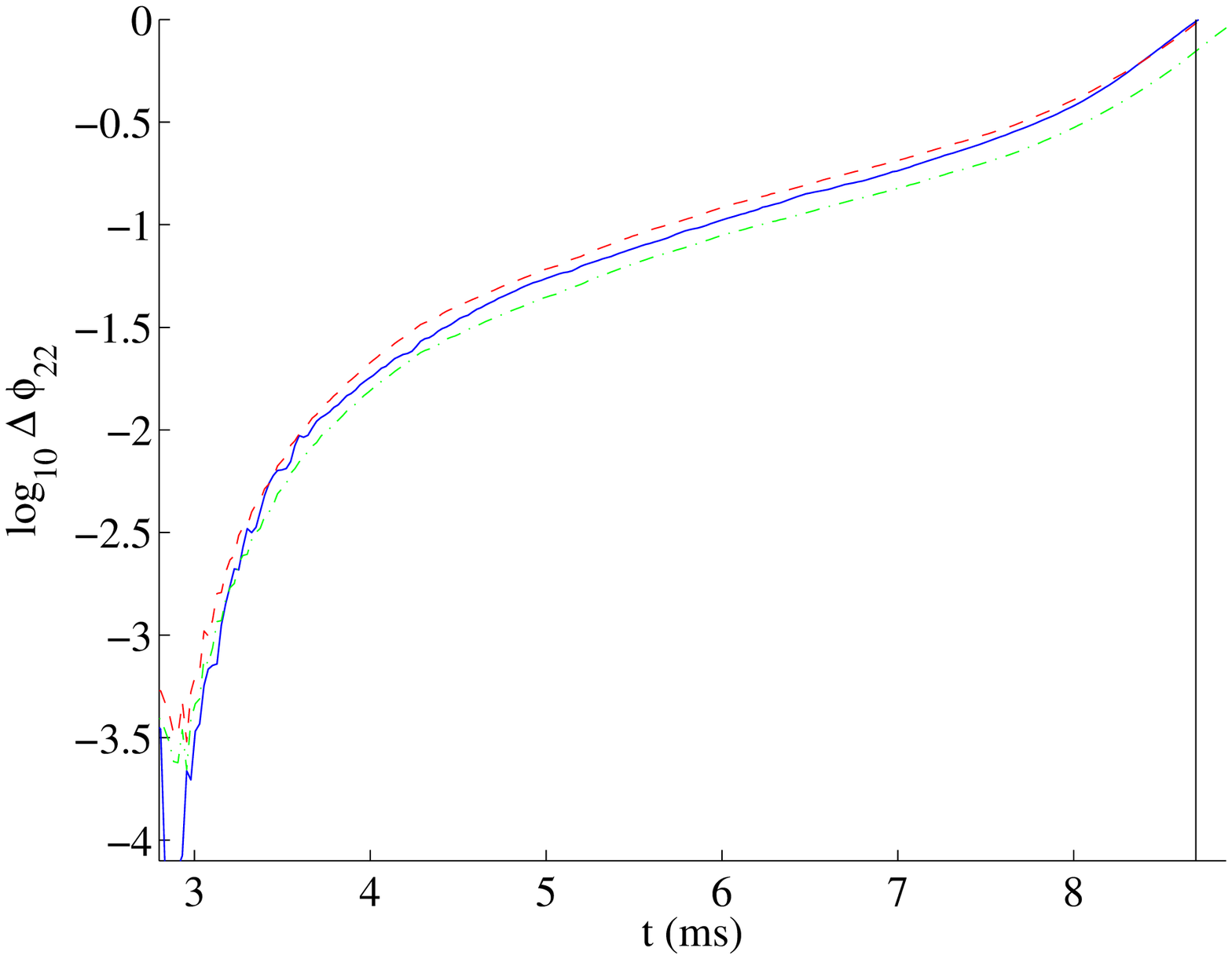}
    \caption{Convergence of phase and
      amplitude in $r\,\psi^4_{22}$ waveform from G2P14 evolution. The
      figure shows the difference between runs H1 and H2 (blue solid
      line), between runs H2 and H3 (green solid-dotted line) as well
      as the difference between H2 and H3 (red dashed line) scaled for
      2nd order convergence. Top panel displays the logarithm of
      amplitude differences, bottom panel displays the logarithm of
      phase differences. The vertical line indicates the merger.
    }
    \label{fig:G2P14:rpsi4_conv} 
  \end{center}
\end{figure}

Fig.~\ref{fig:G2P14:rpsi4_conv} displays the (logarithmic) differences
between the $r\,\psi^4_{22}$ amplitude (top panel) and phase (bottom
panel). The difference between H2 and H3 scaled for second order
convergence is also plotted. The vertical line in the figure marks the
merger time as computed from the waveforms extrapolated in
resolution. The initial junk radiation is also cut out from the figure. 
We observe a quite clear 2nd order convergence in the amplitude while
for the phase the convergence appears slower. The phase error is also
the dominating error in the waveforms at different resolutions.  
A direct inspection of Fig.~\ref{fig:G2P14:rpsi4_diffres} already
highlights this fact. The experimental convergence rate measured for
the phase is a factor between 1 and 2, which suggests that the simulations
have ``just entered'' a convergent regime but more resolution or
higher order numerical methods 
are required. A proper time shift of the waveforms may locally eliminate the
phase error thus improving the results. We prefer not to
perform such procedure in order to keep the analysis simple and clean.
The practice of aligning waveforms 
for convergence tests has been abandoned in some BBH
simulations even in the more complicated cases of unequal masses and
spins, see e.g.~\cite{Pollney:2009yz,Hannam:2010ec,Lousto:2010tb}.

Convergence in the waveforms is lost soon after the merger. In the post-merger phase
a first order in norm convergence is observed in several global quantities, 
cf.~Fig.~\ref{fig:G2P14:Ham} and the discussion in the previous section. 
From our data it is not possible to estimate a precise point-wise convergence rate in the waveforms: 
the convergence in amplitude smoothly drops down to first order, 
the phase shows over-convergence. We observe however a monotonic dependence on 
resolution as evident from Fig.~\ref{fig:G2P14:rpsi4_diffres}. The reason for this behaviour is again 
the truncation error of the HRSC scheme in this strongly nonlinear hydrodynamical phase 
(characterized by a rapid variability in space and time)
which dramatically affects the waveforms.

Using the Richardson method we extrapolate the inspiral waveforms in
order to estimate errors in amplitude and phase. Results concerning
the maximum error estimated during 
the inspiral are reported in Tab.~\ref{tab:gws:errors}. Using the four
resolutions and assuming 2nd order convergence, we obtain $\psi_4$ 
waveforms with a maximum phase error of $\max\delta\phi\sim0.3$~rad
and $\max\delta A/A\sim7$~\%. Assuming 1st order convergence gives instead
$\max\delta\phi\sim1$~rad and $\max\delta A/A\sim24$~\%. Similarly using
only the three highest resolutions gives $\max\delta\phi\sim0.6$~rad
and $\max\delta A/A\sim14$~\% for 2nd order and
$\max\delta\phi\sim1.2$~rad but an unacceptable amplitude error for
the 1st order assumption. 
The errors on the metric waveform are comparable while generically
smaller. 
Furthermore we note that the assumption of 1st order convergence
leads to an evidently non realistic estimate of the merger time from
the extrapolated data $t_{\rm m}=2280$~($11.23$~ms) (H1-3 data) 
while it gives a more reasonable $t_{\rm m}=1800$~($8.87$~ms) for 2nd
order (H1-3 data). 

Some care is needed in interpreting these results because of the
previous discussion. However, from our results we conclude that:
(i)~the series H0-2 while showing convergence is too inaccurate and
not reliable for error estimates; 
(ii)~assuming 1st order convergence for the series H1-3 and
H0-3 is not appropriate and overestimates the actual errors; 
(iii)~a conservative and realistic error estimate is provided by H1-3 data 
assuming 2nd order convergence;
(iv)~the error on the extrapolated data can be estimated as the difference between 
the extrapolation with four resolutions (H0-3) and three resolutions (H1-3), 
e.g.\ $\max\delta\phi\sim0.33$~rad  and $\max\delta A/A\sim7.4$~\% for $r\,\psi^4_{22}$.
As indicated in~\cite{Baiotti:2011am}, the main error on the
waveforms is due, as expected, to resolution rather than finite radius
extraction. Since our results are clearly dominated by truncation
error we did not investigate finite extraction errors but left that
study for future work. 

\begin{table}
  \caption{Error estimates during inspiral for
    extrapolated waveforms. Columns: 
    data used for extrapolation, assumed order of convergence, 
    waveform, 
    maximum absolute error in phase, maximum relative error in phase, 
    maximum relative error in amplitude.}
  \label{tab:gws:errors}
\centering
\begin{tabular}{c|ccccc}
  \hline              
  data & r & waveform & $\max\delta\phi$ [rad] & $\max\delta\phi/\phi$ [\%]& $\max\delta A/A$ [\%]\\
  \hline
  \multirow{4}{*}{H0-3}  
  & \multirow{2}{*}{2} & $r\,\psi^4_{22}$ & 0.29 & 0.68 & 6.89 \\
  & & $r\,h_{22}$ & 0.26  & 0.64 & 2.65 \\
  \cline{2-6}
  & \multirow{2}{*}{1} & $r\,\psi^4_{22}$ & 1.04  & 2.40 & 24.02 \\
  & & $r\,h_{22}$ & 0.87  & 2.17 & 13.31 \\  
  \hline
  \multirow{4}{*}{H1-3}  
  & \multirow{2}{*}{2} & $r\,\psi^4_{22}$ & 0.62 & 1.43 & 14.30 \\
  & & $r\,h_{22}$ & 0.53  & 1.32 &  6.85  \\
  \cline{2-6}
  & \multirow{2}{*}{1} & $r\,\psi^4_{22}$ & 1.20 & 2.85 & $\gtrsim100$ \\
  & & $r\,h_{22}$ & 1.10 & 2.81 & $\gtrsim100$  \\
  \hline
  \hline                                   
\end{tabular}
\end{table}


\section{Conclusions}

In this work we presented detailed tests and first full scale
evolutions for a new computer code aimed at 3+1 numerical studies of
general relativistic matter spacetimes. The implementation represents an 
upgrade of the {\bam} code developed previously for vacuum
systems~\cite{Brugmann:2008zz,Bruegmann:2003aw,Bruegmann:1997uc}.
The code solves the flux-conservative formulation of the GRHD
equations~\cite{Banyuls:1997zz} coupled with the BSSNOK system. Mesh
refinement (moving boxes) and a metric solver  
were already provided by the {\bam} code. The GRHD equations are solved by
means of robust HRSC
methods~\cite{DelZanna:2007pk,Kurganov:2000,Nessyahu:1990,Shu:1988,Shu:1989} 
and share the same grid and the same time stepping algorithm with the
metric solver. We described in detail the equations and the numerical
method implemented. The code allows the use of one-parameter EoS tables
and implements the hybridization procedure of~\cite{Shibata:2005ss} to
model thermal effects. We proposed a simple, thermodynamically consistent
interpolation scheme for one-parameter EoSs. 

We validated the code against a number of stringent tests involving
single star spacetimes.  We studied the performance of different
reconstruction procedures and the convergence of the code.
We found that in our set up the use of CENO reconstruction shows a
clear 2nd order convergence  and leads to smaller global truncation errors
compared to the other methods. 

We presented test evolutions of irrotational equal-mass binary
neutron star configurations.  
Matter is described by simple polytropic and ideal gas EoSs.
In both evolutions the formation of an HMNS is observed. 
In the isentropic case it collapses quite rapidly producing a Kerr BH
surrounded by a disk of mass $M_{\rm d}\lesssim 2\times10^{-2}M_0$
while in the other case the HMNS survives for $9$ ms due to
thermal pressure support. 

We investigated the impact of different reconstruction
methods on the dynamics of the HMNS. Our results highlight a strong
influence of the numerical method and of the resolution on the
simulated physics. Precise quantitative statements on the post-merger
phase require extreme care. 

We presented new results concerning the use of different values
of the damping parameter $\eta$ in the shift condition. Small
values of $\eta$ are found to produce a less coordinate-related eccentricity 
during inspiral and to reduce the coordinate size of the final BH.

The gravitational radiation emitted by the system was investigated. 
We characterized the GWs (metric waveforms) and discussed two different
methods to compute them from $\psi_4$ waveform: a time domain (CTI,
e.g.~\cite{Damour:2008te,Baiotti:2008nf}) and Fourier domain
(FFI,~\cite{Reisswig:2010di}) integration. The FFI provides better
results minimizing unphysical drifts and oscillations. 
We performed self-convergence tests of the waveforms.
While the resolutions employed are probably close but not still not
optimal for production  runs, we found 2nd order convergence during 
the inspiral phase without any time-shifting procedure.
For the first time in BNS simulations we consistently estimated
error-bars on the waveforms by extrapolating the results in
resolution.   

This work represents our first contribution to the study of GWs from BNS systems.
The methodology presented here will be applied in future works. 
In particular, we plan in the near future the production of accurate and convergent inspiral waveforms 
from initial configurations described by different realistic EoS in order to investigate the impact of the EoS on the GW signal~\cite{Read:2009yp,Hinderer:2009ca,Baiotti:2010xh,Baiotti:2011am}.


\begin{acknowledgments}
  The authors sincerely thank David Hilditch and Alessandro Nagar 
  for many valuable discussions. The authors thank Doreen M\"uller
  for discussion on metric waveform integration.
  The authors thank the Meudon group for making publicly available 
  \texttt{LORENE} initial data and Eric Gourgoulhon for explanations. 
  During the development of the {\bam} matter code and of this paper
  the authors benefitted from valuable discussions and correspondence
  with Luca Baiotti, Thibault Damour, Roman Gold, Roberto De Pietri, Harald
  Dimmelmeier, Luca Del Zanna, Frank L\"offler, Pedro Montero, Luciano
  Rezzolla, and Wolfgang Tichy.
  This work was supported in part by  
  DFG grant SFB/Transregio~7 ``Gravitational Wave Astronomy''.
  Computations where performed at LRZ (Munich) and on local
  clusters at TPI (Jena).
\end{acknowledgments}

\appendix
\section{Algorithm to recover primitives}
\label{app:c2p}

In this appendix we summarize the procedure and the equations to
recover the primitive variables from the conservative
ones. The specific algorithm adopted has been developed in a number of previous publications~\cite{Marti:1991wi,Marti:1994,Marti:1996,Dimmelmeier:2002bk,Font:1998hf,Baiotti:2004wn}. 
Other algorithms have been developed, e.g.~\cite{Dean:1994,Dolezal:1995}, 
in particular we have successfully tested the method of~\cite{Dean:1994} in single star evolutions. 
We leave for future work extensive tests and comparisons between different algorithms.

We first discuss an iterative procedure for a general EoS,
$p=P(\rho,\epsilon)$, first employed in~\cite{Marti:1991wi}.
Specific procedures can be designed once a specific form of EoS is
given~\cite{Marti:1999wi}. We then present the procedure we adopt in
the case of cold EoSs which is based on an iterative algorithm for
$\rho$~\cite{Baiotti:2004wn}. 
Finally we describe the modifications introduced to handle the
presence of the artificial atmosphere.

Following~\cite{Marti:1994} we invert Eqs.~\eqref{eq:hydro_cons} to find
\begin{eqnarray}
  \label{eq:cons2prim:vi}
  v^i(p) &=& \frac{S^i}{\tau + D + p} \ , \\
  \label{eq:cons2prim:W}
  W(p) &=& \frac{\tau+p+D}
  { \sqrt{ \left( \tau+p+D \right)^2 - S^2 } } \ , \\
  \label{eq:cons2prim:rho}
  \rho(p) &=& \frac{D}{W} \ ,  \\
 \label{eq:cons2prim:eps:bis}
  \epsilon(p) &=&
  D^{-1} \left[  \sqrt{ \left( \tau+p+D \right)^2 - S^2 } - W p - D\right].
\end{eqnarray}
Primitive variables can be computed from the conservative variables using the
above relations once the pressure is known.
The pressure is determined by the EoS looking for the root of the
nonlinear algebraic equation
\begin{equation}
  \label{eq:f_of_p}
  f(p) \equiv p - P\left( \rho(p),\epsilon(p)\right) = 0 \ .
\end{equation} 
The algorithm used is the Newton-Raphson iteration,
\begin{equation}
  \label{eq:NewRap}
  p^{\rm new} = p^{\rm old} - \frac{f(p)}{f'(p)} \ .
\end{equation} 
The derivative of $f$ is given by
\begin{eqnarray}
  \label{eq:fx_of_p}
  f'(p) &=& 1 - \chi \frac{\partial \rho}{\partial p} 
  - \kappa \frac{\partial \epsilon}{\partial p} \ ,\\
  \label{eq:drhodp}
  \frac{\partial \rho}{\partial p} &=&
  \frac{D S^2}{(D+p+\tau )^2 \sqrt{(D+p+\tau )^2-S^2}} \ , \\
  \label{eq:depsldp}
  \frac{\partial \epsilon}{\partial p} &=&
  \frac{p S^2}{D \left((D+p+\tau )^2-S^2\right)^{3/2}} \ .
\end{eqnarray} 
In the case of a one-parameter EoS, $p=P(\rho)$, $h=h(\rho)$,
$\epsilon=\epsilon(\rho)$, and 
\begin{equation}
  W = \sqrt{1+\frac{S^2}{(Dh)^2}}
\end{equation}
are functions of the density and the conservative variables only.
Primitive variables can be computed from $\rho$, once the latter
is determined by
\begin{equation}
  \label{eq:g_of_rho}
  g(p) = W(\rho) \rho - D \ ,
\end{equation} 
using again a Newton-Raphson root finder with
\begin{eqnarray}
  \label{eq:gx_of_rho}
  g'(\rho) &=& W(\rho) - \rho\frac{S^2 h'(\rho)}{W D^2 h^3} \ , \\
  h'(\rho) &=& \epsilon'(\rho) - \frac{P}{\rho^2}  + \frac{\chi}{\rho}
  \ .
\end{eqnarray}
Note that $h'(\rho) =\frac{\chi}{\rho}$ if the principle of
thermodynamics at zero temperature is applied.

The recovery procedure for a general EoS can fail at low densities, 
e.g.\ in presence of the atmosphere.
A reason for this is simply machine accuracy: since typically $p\ll
D$, for very low densities the Newton-Raphson algorithm does not
converge. A further complication is the spuriously high value of the
velocity generated by the artifical atmosphere treatment. 
To handle these problems, similarly
to~\cite{Font:1998hf,Baiotti:2004wn}, we combine both the algorithms
described above and we implement a set of hierarchic prescriptions which
are found to work in practice. Specifically: 
(i)~every time a grid point reaches a density below the atmosphere
threshold density the code sets atmosphere values both in
primitives and conservatives and continues to the next point;
(ii)~if the general EoS algorithm does not converge due to a too small
value of the pressure, i.e.\ $p^{\rm new}-p^{\rm old}<\varepsilon\sim
10^{-12}$, then atmosphere values are set; 
(iii)~if it returns unphysical values for $\epsilon,\rho$ or $p$ then
the code tries the inversion with the algorithm for the cold EoS;
(iv)~if it returns unphysical values of $v^2$ then atmosphere values
are set; 
(v)~if the algorithm for the cold EoS does not converge or returns
unphysical values not cured by finer grid levels, then the code stops.
In practice we observe the necessity of (ii), (iii) and (iv) only in
binary simulations.


\bibliographystyle{apsrev}     
\bibliography{refs20110727}{}

\end{document}